\documentclass[preprint,showpacs,preprintnumbers,amsmath,amssymb]{revtex4}

\usepackage{graphicx}
\usepackage{dcolumn}
\usepackage{bm}
\usepackage{amsmath,dsfont}
\usepackage{amssymb,amsthm,amscd, amsbsy, array}
\usepackage{graphics,graphicx,xcolor}
\usepackage{etex}
\usepackage[all]{xy}
\input amssym.def
\input amssym.tex
\usepackage{color}
\newcommand{\red}{\textcolor{red}}
\newcommand{\blue}{\textcolor{blue}}

\renewcommand\thesection{\arabic{section}}
\numberwithin{equation}{section}

\newcommand{\half}{{\scriptstyle{\frac{1}{2}}}}
\def\2{{\half}}
\newcommand{\const}{\mathop{\rm const}\nolimits}
\def\parag{\hfil\break} 
\def\kikezd{\parag\underbar}
\def\su{{\rm su}}
\def\p{{\partial}}

\def\vPi{{\bm{\Pi}}}

\newcommand{\vP}{{\bm P}}
\newcommand{\vQ}{{\bm Q}}

\newcommand{\vp}{{\bm p}}
\def\vE{{\bm E}}

\def\beq{\begin{equation}}
\def\eeq{\end{equation}}
\def\beqa{\begin{eqnarray}}
\def\eeqa{\end{eqnarray}}
\def\nn{\nonumber}
\def\barray{\left(\begin{array}}
\def\earray{\end{array}\right)}
\def\barraynb{\begin{array}}
\def\earraynb{\end{array}}
\def\ort{{\rm o}}
\def\Ort{{\rm O}}

\def\so{{\rm so}}
\def\SO{{\rm SO}}
\def\smallover#1/#2{\hbox{$\textstyle\frac{#1}{#2}$}} %

\def\vx{{\bm{x}}}
\def\vy{{\bm{y}}}

\def\vX{{\bm{X}}}

\def\vPi{{\bm{\Pi}}}

\newcommand{\vK}{{\bm K}}

\usepackage{color}
\def\blue#1{{\color{blue}#1}}
\def\red#1{{\color{red}#1}}
\def\green#1{{\color{green}#1}}

\begin{document}

\preprint{arXiv:1207.2875v7}

\title{Newton-Hooke type symmetry of anisotropic oscillators \footnote
{Dedicated to the memory of J.-M. Souriau, deceased on March 15 2012, at the age of 90.}}

\author{P. M. Zhang$^{1}$\footnote{email:zhpm@impcas.ac.cn},
P. A. Horvathy$^{1,2}$\footnote{email:horvathy@lmpt.univ-tours.fr},
K. Andrzejewski$^{3}$\footnote{email:k-andrzejewski@uni.lodz.pl},
J. Gonera$^{3}$\footnote{email:jgonera@uni.lodz.pl},
P. Kosi\'nski$^{3}$\footnote{email: pkosinsk@uni.lodz.pl}
}
\affiliation{$^1$Institute of Modern Physics,
Chinese Academy of Sciences
\\
Lanzhou, China
\\
$^2$Laboratoire de Math\'ematiques et de Physique
Th\'eorique,
Universit\'e de Tours, France
\\
$^3$Faculty of Physics and Applied Informatics, University of Lodz, Poland,
}

\date{\today}

\begin{abstract}
The rotation-less Newton--Hooke - type symmetry found recently in the
Hill problem and instrumental for explaining the center-of-mass decomposition
is generalized to an arbitrary anisotropic oscillator in the plane.
Conversely, the latter system is shown, by the orbit method, to be the most
general one with such a symmetry. Full Newton-Hooke symmetry is recovered in the isotropic case. Star escape from a Galaxy is studied
as application.
\end{abstract}

\pacs{
11.30.-j,   
02.40.Yy,   
02.20.Sv,   
\\[8pt]
%
%
}

\maketitle

\tableofcontents

\newpage

\section{Introduction}\label{Intro}

Renewed interest in Kohn's theorem on decomposing a system of charged particles
in a magnetic field into center-of-mass and relative coordinates stems from relating
it to the Newton-Hooke symmetry of the Landau problem
\cite{GiPo, NHlit, ZH-Kohn,ZH-Kohn-II,AGK}.

Yet another example was found recently, however,  namely for  Hill's equations
\cite{ZGH,ZH-Hill}. Remember that Hill originally devised his method for finding
approximate solutions of the three-body problem, and in particular for the Moon-Earth-Sun
system \cite{HillAJM,Gutzwiller}. Further applications involve stellar dynamics
\cite{Bok,Mineur,Heggie,Heggie2}.

Hill's equations also admit a center-of-mass decomposition, but no full Newton-Hooke
symmetry. Rotations are broken, but translations and (generalized) boosts are still symmetries,
hinting at that it is the subgroup spanned by the latter which is important for the purpose;
additional symmetries like rotations are secondary.

In this paper, when referring to \emph{Newton-Hooke type} symmetry it is
\emph{Newton-Hooke with or without rotations} that we have in mind.

The possibility of decomposing an \emph{isolated system} into center-of-mass and
relative coordinates has been linked to Galilean symmetry~: Souriau \cite{SSD} argued,
in fact, that this property depends on the Galilei group having an invariant
Abelian subgroup --- namely the one spanned by \emph{translations} and (Galilean)
\emph{boosts}.

Remarkably, the cohomological structure which determines the existence of a central
extension originates precisely in this subgroup \cite{Chevalley,SSD}. Remember that,
in dimension $d\geq3$, both the Newton-Hooke and the Galilei groups have a one-parameter
central extension, but in the plane they both admit an ``exotic'', \emph{two-parameter}
central extension \cite{LL,centralex,DHexo,NHcoho}.

In this paper, we focus our attention at the Newton-Hooke case, the Galilean one being
rather well-known.

The ordinary Landau problem admits the one-parameter centrally extended Newton-Hooke
group as symmetry \cite{GiPo}, but the ``exotic'' [non-commutative] version has indeed
the two-parameter version \cite{ZH-Kohn-II}. In the Hill case, rotation-less
``Newton-Hooke type'' symmetry, with one (or, in the ``exotic'' case, with two)
central extensions could be established \cite{ZGH,ZH-Hill}.

Then Our main result, proved in Section \ref{orbitconst}, says~:

\vskip7mm
\goodbreak
\kikezd{Theorem~1}~: \textit{The most general planar system
with Newton--Hooke - type symmetry  is a (possibly non-commutative) anisotropic oscillator
in a uniform magnetic background. The symmetry extends
to full Newton-Hooke symmetry in the isotropic case.}
\goodbreak

\vskip3mm
The proof will be accomplished by applying the \emph{orbit method}, which provides
us indeed with all systems upon which the symmetry group acts transitively
\cite{SSD,Kirillov,Arnold,Giac}.

From the technical point of view, we will find it convenient to use \emph{chiral decomposition}
\cite{Banerjee,AGKP,NHcoho,ZH-chiral,ZGH,ZH-Hill}. The motion in the (ordinary) Hall effect
can in fact be decomposed into two uncoupled chiral oscillators with opposite chirality
\cite{Banerjee}. Conversely, combining two $1$d chiral oscillators may yield the non-commutative
Landau problem \cite{AGKP, Sochi,NCLandau,ZH-chiral}; then the chiral method allows for
an elegant derivation of the (Newton-Hooke) symmetry.

Recently, the method was extended to the Hill problem \cite{ZGH,ZH-Hill} which is in fact
a ``maximally anisotropic oscillator''; here we further extend it to arbitrary anisotropy.

Our paper is organized as follows.

In section \ref{chiralosc} chiral oscillators are reviewed.

Then, after outlining the Landau problem and its Newton-Hooke symmetry, we turn to the
Hill problem. After some remark on its application to stellar dynamics, we present its
 the rotation-less Newton--Hooke type symmetry.

In Section \ref{anisotropic}, we generalize to an arbitrary, possibly anisotropic, oscillator.

In Section \ref{orbitconst} we proceed conversely: applying the orbit method we describe all
systems with Newton-Hooke type symmetry acting transitively.

We also study the arising of further symmetries and explain the difference between symmetry
with or without rotations. Our results allow us to deduce~:

\kikezd{Theorem~2}~: \textit{The system is either a truly anisotropic oscillator with
Newton-Hooke-type symmetry only and no rotations, or it is isotropic with full Newton-Hooke
symmetry, including rotations.}

\vskip3mm
Moreover, in the first case, it can be brought into a ``Hill-type form'', and in the second
one it can be transformed into a free particle, cf. Sections \ref{further} and \ref{Bargmann}.

\vskip2mm
An outlook is presented in the Conclusion, section \ref{Concl}.

\section{Chiral oscillators}\label{chiralosc}

Chiral oscillators arise owing to the ambiguity of the phase-space description of a harmonic
oscillator. In detail, let us consider a one-dimensional harmonic oscillator of unit mass
$m=1$ and frequency $\omega$. Viewing the position and velocity, $x$ and $\dot{x}$, simply
as coordinates on the phase space, we write
\beq
y_1=x,
\qquad
y_2=\dot{x},
\label{phaseSC}
\eeq
and consider the two first-order phase space Lagrangians
\beq
L_\pm=\pm\frac{1}{2}\epsilon_{ij}y_i\dot{y}_j-\frac{\omega}{2}\,\vy^2\,.
\label{chiralLag}
\eeq
The associated (Euler-Lagrange) equations read
\beq
\dot{y}_i=\mp\omega\,\epsilon_{ij}y_j.
\label{chireqmot}
\eeq
Our clue is that, for both signs in eqn. (\ref{chiralLag}), eliminating either $y_1$ or
$y_2$ yields, for the remaining variable, the \emph{same} equation, namely that of a $1$d
harmonic oscillator,
\beq
\ddot{y}_i+\omega^2y_i=0,
\qquad
i=1,2.
\eeq

\begin{figure}
\begin{center}
\includegraphics[scale=.5]{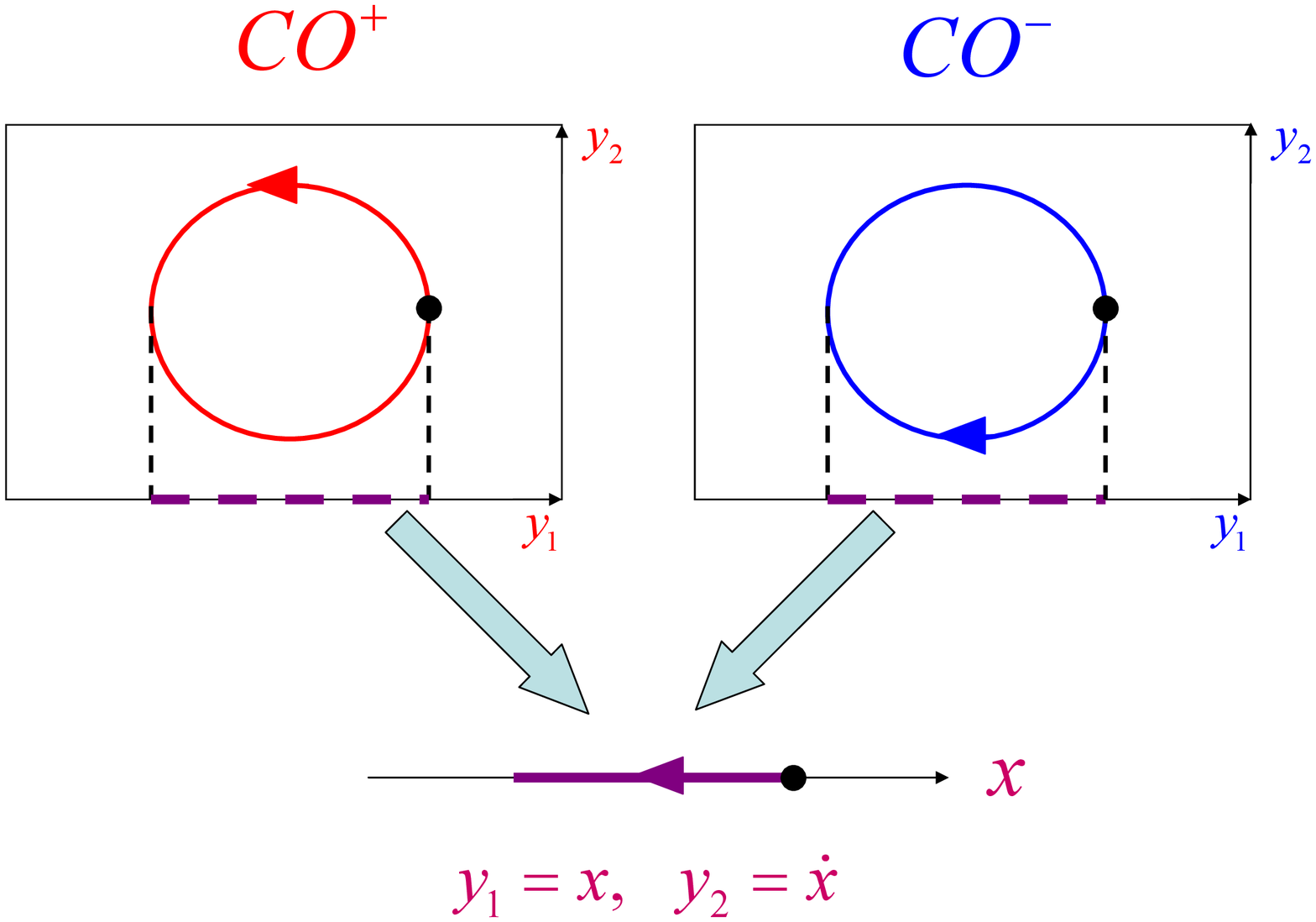}
\vspace{-8mm}
\end{center}
\caption{\it The phase-space trajectory of a chiral oscillator turns clockwise or
anti-clockwise, depending on the sign of the frequency. Both trajectories project,
however, onto the \emph{same} motion in configuration space.
}
\label{pic-chiral}
\end{figure}

The solutions of (\ref{chireqmot}) are simple rotations in phase space -- but in
\emph{opposite} directions, depending on the sign \footnote{This is particularly clear
if we use the equivalent Lagrangian
$$
\widetilde{L}_\pm=\frac{1}{2}\epsilon_{ij}y_i\dot{y}_j\mp\frac{\omega}{2}\,y_i^2\,.
\label{chiralLagbis}
$$}.
(This is indeed the very meaning of the word ``chiral''). Then we note that \emph{both}
types of motions project into configuration space according to the \emph{same} motion
$x(t)$, as illustrated on Fig. \ref{pic-chiral}.

We note that the same conclusion can be reached using a Hamiltonian framework. The eqns.
(\ref{chireqmot}) are indeed those of the symplectic structure and Hamiltonian
\beq
\Omega_\pm=\pm\half\epsilon_{ij}dy_i\wedge dy_j,
\qquad
H=\half\omega \vy^2,
\label{chirsympHam}
\eeq
namely
$\dot{y}_i=\big\{y_i,H\big\}_\pm$, where the Poisson brackets $\big\{\,\cdot\,,\,\cdot\big\}_\pm$
are those  associated with the chosen symplectic structure $\Omega_\pm$. The coordinates $y_i$ are
non-commuting,
\beq
\big\{y_1,y_2\big\}_\pm=\mp 1,
\eeq
--- as it is natural for position and momentum on the phase space. We mention for completeness
that the Lagrangians (\ref{chiralLag}) are the Cartan forms of the Souriau forms \cite{SSD,APH},
\beq
{L_\pm}dt=\lambda_{\pm},
\qquad
d\lambda_{\pm}=\Omega_{\pm}-dH\wedge dt.
\eeq

\section{Landau problem}\label{Landau}

The classical example of a system with one-parameter-centrally-extended Newton-Hooke
symmetry is provided by the ``ordinary'' [meaning commutative] Landau problem \cite{GiPo}.
Generalizing the latter, we consider $N$ ``exotic'' particles endowed with masses,
charges and non-commutative parameters $m_a,\,e_a$ and $\theta_a$, respectively,
moving in a planar electromagnetic field $B, \vE$ \cite{ZH-Kohn-II}.
Following  \cite{DHexo,NCLandau}, we describe our system by
\beq\barraynb{lll}
m^*_a
\dot{x}^i_a&=&{p}^i_a-m_ae_a\theta_a\varepsilon^{ij}E^j,
\\[6pt]
\dot{{p}}^i_a&=&e_aB\varepsilon^{ij}\dot{x}^j_a
+e_aE^i,
\earraynb
\label{exoNeqmot}
\eeq
where
\beq
m^*_a=\Delta_am_a
\quad\hbox{ with}\quad
\Delta_a=1-e_a\theta_a B
\eeq
is the effective mass of the particle labeled by $a=1,\dots,N$. Note, in the first relations,
 also  the ``anomalous velocity terms'' perpendicular to the electric field, $\vE$. The
 variables $\vp_a$ here \emph{could} be called ``momenta'' -- but to avoid confusion with
 the conserved quantities, we simply consider them as coordinates on the phase space.

Although our theory works for any $B$ and $\vE$, we assume, for simplicity, that the
magnetic field is constant, $B=\const$, and that the electric field is that of an
isotropic harmonic trap, $-{k}\vx$, augmented with an interparticle force coming from
some two-body potential, $V=\sum_{a\neq b}V_{ab}(\vx_a-\vx_b)$.

For $\theta_a=0$, the ordinary Landau problem is recovered.

Summing over all particles, we find that when ${e_a}/{m_a}$ and $e_a\theta_a$ are both
constants i.e. when the \emph{generalized Kohn conditions} \cite{ZH-Kohn-II},
\beq
\kappa_a\equiv\frac{e_a}{m_a}=\frac{e}{M}\equiv\kappa,
\qquad
e_a\theta_a=e\,\Theta,
\qquad
\Theta=\frac{\sum_am_a^2\theta_a}{M^2}
\label{exoKohnCond}
\eeq
hold (where $M=\sum_am_a$ and $e=\sum_ae_a$ are the total mass and charge), then the
center-of-mass, $\vX=\sum_a{m}_a\vx_a/M$, splits off,
\beq\barraynb{lll}
M^*\dot{X}^i&=&P^i-M{e}\Theta\varepsilon^{ij}E^j,
\\[4pt]
\dot{P}^i&=&eB\varepsilon^{ij}\dot{X}^j+
eE^i,
\earraynb
\label{exoCMeqmot}
\eeq
where
\beq
M^*=\Delta M,\qquad
\Delta=1-{e}\Theta B,
\qquad
\vP=\sum_a\vp_a.
\eeq
The center-of-mass behaves hence as a \emph{single ``exotic'' particle} carrying the
total mass, charge and non-commutative parameter, $M,\,e$ and $\Theta$, respectively.

We note that eqns. (\ref{exoCMeqmot}) are in fact Hamilton's equations for the Poisson
structure \cite{NCLandau},
\beqa
H&=&\frac{\vP^2}{2M}+V(\vX)\,,
\label{Ham}
\\[8pt]
\{X^i,X^j\}&=&\frac{\Theta}{\Delta}\,\varepsilon^{ij},
\qquad
\{X^i,P^j\}=\frac{\delta^{ij}}{\Delta},
\qquad
\{P^i,P^j\}=\frac{eB}{\Delta}\,\varepsilon^{ij}.
\label{exoPB}
\eeqa

\begin{figure}
\begin{center}
\includegraphics[scale=1]{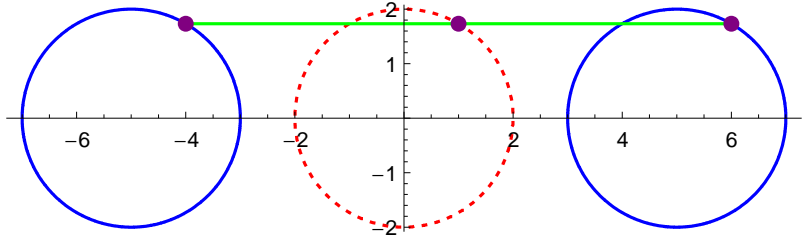}
\vspace{-8mm}
\end{center}
\caption{\it If the Kohn conditions (\ref{exoKohnCond}) are satisfied, all particles
turn along circles of equal radii with common angular velocity. Their motion is
shared by their center of mass (in dashed).
}
\label{KohnCFig}
\end{figure}

When the [generalized] Kohn condition (\ref{exoKohnCond}) is satisfied, then, for
identical initial velocities, all individual particles move in the same way, ---
and this motion is shared by their center of mass, cf. Fig. \ref{KohnCFig}.

\begin{figure}
\begin{center}
\includegraphics[scale=1]{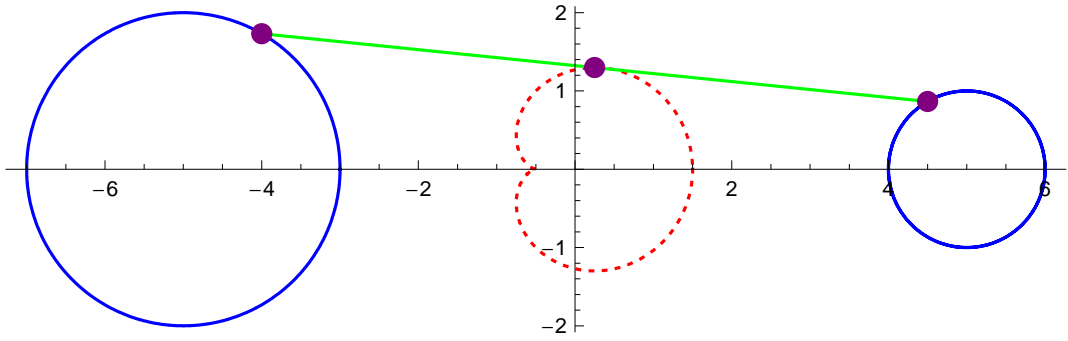}
\vspace{-8mm}
\end{center}
\caption{\it If the Kohn conditions (\ref{exoKohnCond}) are \underline{not} satisfied,
$\kappa_2\equiv e_2/m_2=2e_1/m_1\equiv2\kappa_1$, for example, the individual radii and
the angular velocities are different. The motion is not more collective, and the
center of mass describes a complicated (dashed) curve.
}
\label{noKohnCFig}
\end{figure}

\begin{figure}
\begin{center}
\includegraphics[scale=.36]{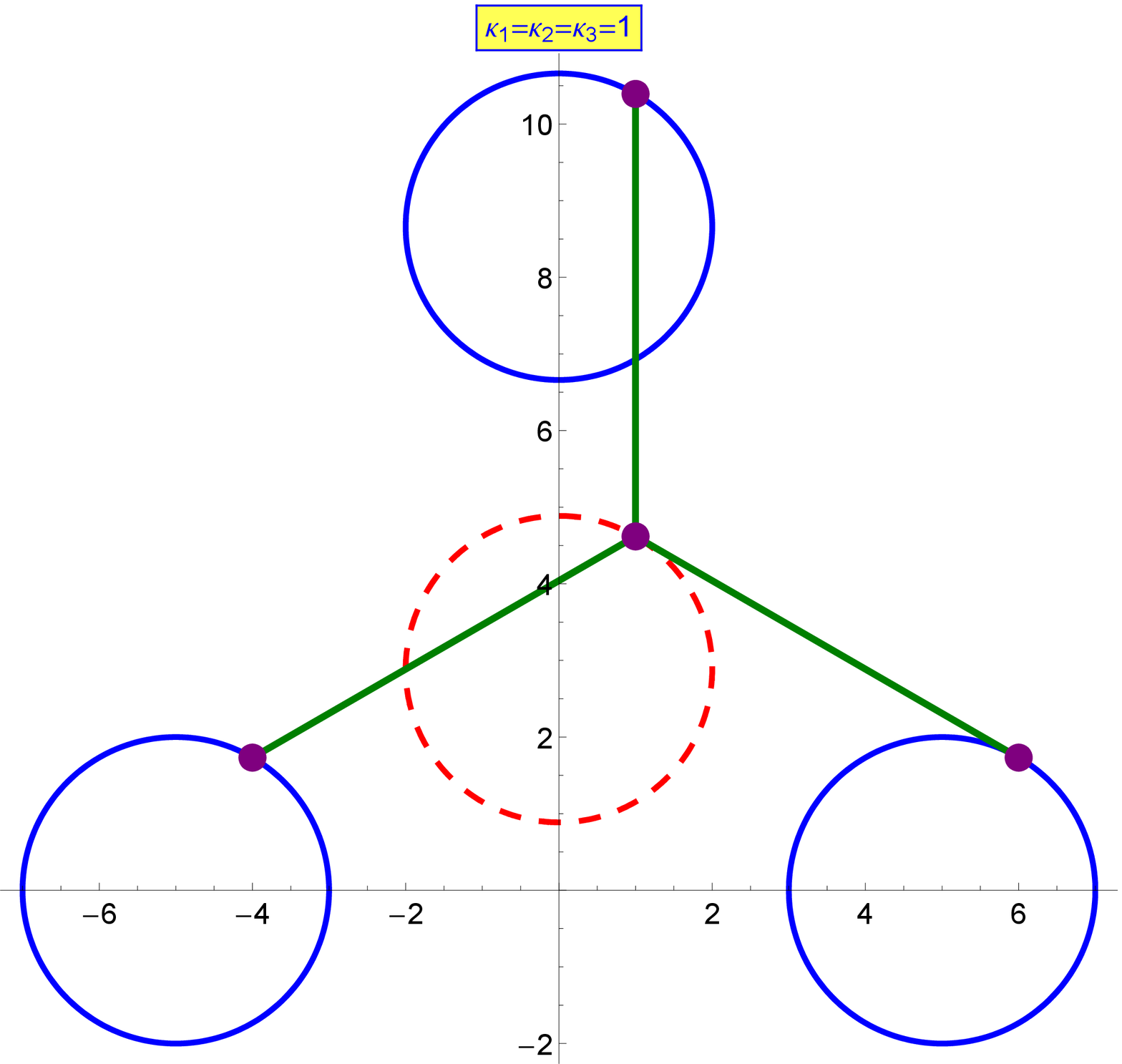}\;
\includegraphics[scale=.39]{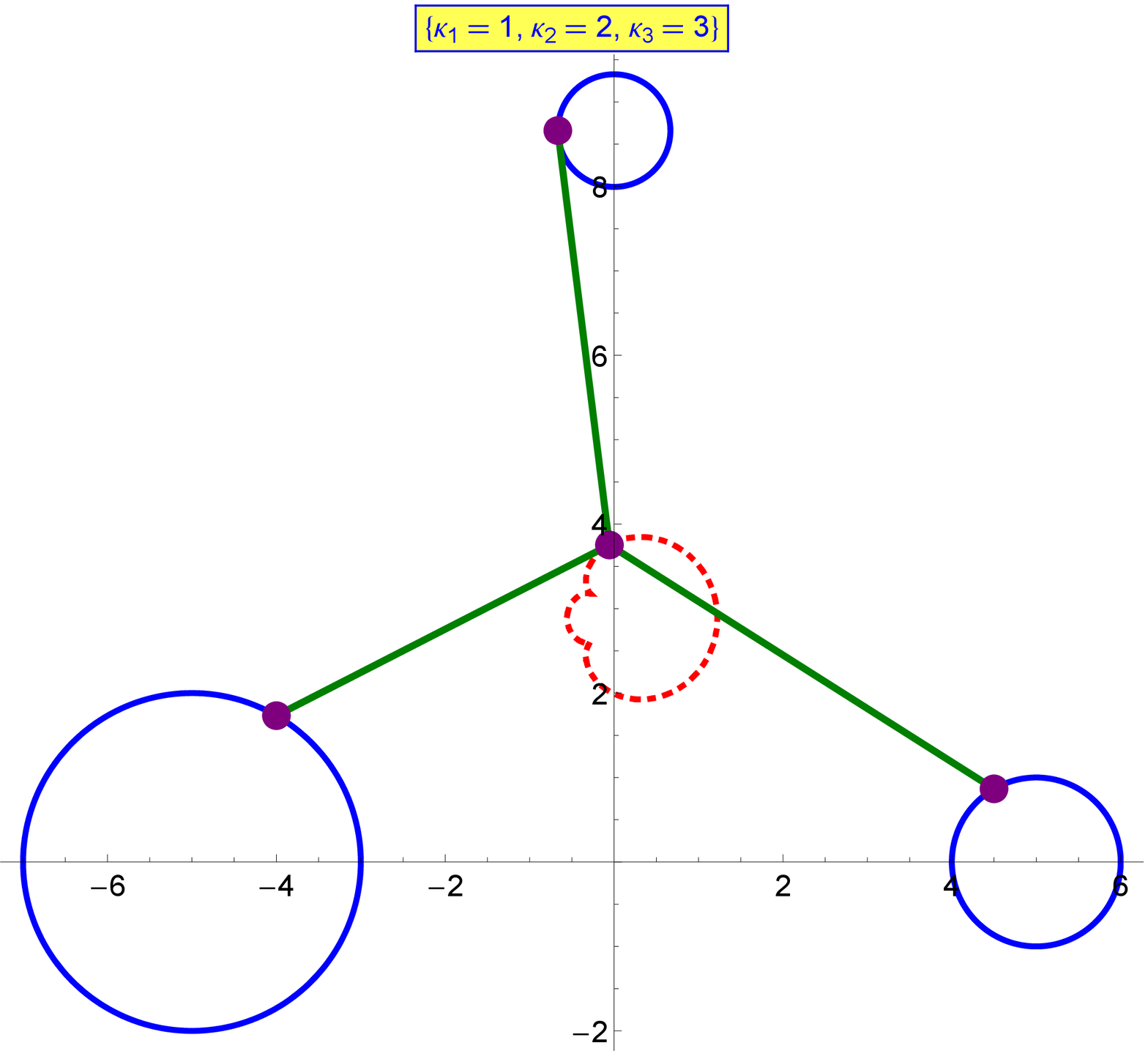}
\vspace{-8mm}
\end{center}
\caption{\it The behavior of a 3-body system. (a) If the Kohn conditions
(\ref{exoKohnCond}) are satisfied, all particles move collectively, along with
their center-of-mass. (b) If (\ref{exoKohnCond}) is \underline{not} satisfied,
$\kappa_1=1, \kappa_2=2,\kappa_3=3$, for example, the motion is not more collective,
and the center of mass (in dashed) describes a complicated curve.}
\label{kohn3Figs}
\end{figure}

The best way to understand the intuitive content of the Kohn condition is, however,
to consider what happens when the Kohn condition is \emph{not} satisfied. Consider,
for example, two particles in a pure magnetic field such that
\beq
\kappa_2\equiv\frac{e_2}{m_2}=2\kappa_1\equiv2\,\frac{e_1}{m_1}\,.
\eeq
Then, assuming identical initial velocities, each of them performs a rotational motion
but with \emph{different radii},
\beq
R=(m/e)\,\frac{v}{B}
\qquad\Rightarrow\qquad
R_2=\frac{1}{2}R_1,
\eeq
and with \emph{different frequencies}:
\beq
\omega=\frac{v}{R}\quad\Rightarrow\quad
\omega_2=2\omega_1
\eeq
[so that $\omega_1R_1=v=\omega_2R_2$].
Their center-of-mass would then clearly \emph{not} move on a circle, rather on some
complicated curve, cf. Fig. \ref{noKohnCFig}. The 3-body situation is illustrated
on Fig. \ref{kohn3Figs}.

\kikezd{Symmetries}.

Let us restrict ourselves henceforth to the purely magnetic case, $\vE=0$ and to the
center-of-mass motion. The coordinate $\vP$ is not conserved; one readily shows, however,
that the ``magnetic momentum'' (which can also be derived by Noether's theorem as the
conserved quantity associated with the translational symmetry \cite{ZH-chiral})
\footnote{For the record, $\vPi$ and $\vP$ are related as $P^i =\vPi^i+M \omega\varepsilon^{ij} X^j$.}
and ``magnetic center-of-mass'',
\beq\barraynb{lll}
\Pi_i&=&M\Delta\big(\dot{X_i}-\omega^*\varepsilon_{ij}X_j\big),
\\[6pt]
\vK&=&M\Delta^2 R(\omega^*t)\dot{\vX},
\earraynb
\label{magmomboost}
\eeq
respectively, where  $\omega^*=eB/M^*=\omega/\Delta$, are both conserved \footnote{Note that
$dK^i/dt=\p_tK^i+\{H,K^i\}=0$ as it should.}, and span indeed two uncoupled Heisenberg
algebras with central charges $-M\omega$ and  $\Delta M\omega$,
\beqa
\{\Pi^i,\Pi^j\}=-M\omega\varepsilon^{ij},
\qquad
\{K^i,K^j\}=\Delta M\omega\varepsilon^{ij},
\qquad
\{\Pi^i,K^j\}=0.
\label{exoCMcomrel}
\eeqa
Time translations and rotations  are plainly symmetries also, and the associated conserved
quantities, namely the Hamiltonian $H$ in (\ref{Ham}) \footnote{An isotropic harmonic
electric force can be freely added, cf. \cite{ZH-chiral})}, augmented with the total
angular momentum \cite{DHexo,NCLandau},
\beqa
J=\vX\times\vP+\frac{eB}{2}\vx^2+\frac{\Theta}{2}\vP^2,
\label{totangmom}
\eeqa
have commutation relations
\beqa
\big\{H,\Pi^i\big\}&=&0,
\qquad
\big\{H, K^i\big\}=
-\frac{\omega}{\Delta}\,\varepsilon^{ij}K^j,
\\[8pt]
\big\{J,X^i\big\}&=&\varepsilon^{ij}X^j,
\qquad
\big\{J,P^i\big\}=\varepsilon^{ij}P^j,
\qquad
\big\{J,H\big\}=0.
\eeqa

In conclusion, the exotic Landau problem [with or without an isotropic harmonic trapping
force] admits an ``exotic'' i.e. two-parameter centrally extended Newton-Hooke symmetry
\cite{ZH-Kohn-II,ZH-chiral}. In the commutative case $\Theta=0$, the  central charges are
correlated, $\mp M\omega$, and the symmetry reduces to the one-parameter extension studied
in \cite{GiPo}.

We record for further use that the total angular momentum, $J$ in (\ref{totangmom}), can
also be presented in a number of different ways. Firstly, we note that the new variables
\cite{DHexo}%
\beqa
Q_i&=&x_i+\frac{1}{eB}\left(1-\sqrt{1-\theta eB}\right) \varepsilon_{ij}p_j\,,
\\[8pt]
P_i&=&\frac{1+\sqrt{1-\theta eB}}2p_i-\frac{1}{2eB}\varepsilon_{ij}x_j\,,
\eeqa
are canonical, and in their terms the total angular momentum is simply
\beq
J=\vQ\times\vP.
\label{QPangmom}
\eeq
Here we just mention that using chiral coordinates (sect. \ref{chiralsymm}), the angular
momentum can further  be decomposed, see (\ref{chiralangmom}).

From now on the generalized Kohn conditions (\ref{exoKohnCond}) will always be assumed,
allowing us to consider the center-of-mass alone. Coordinates will again be denoted by
lower-case letters, as for a one-particle theory.

\bigskip
\section{The Hill problem}\label{HillSection}

Hill's original aim has been to study  the Moon-Earth-Sun system \cite{HillAJM,Gutzwiller}.
Later, his technique has been applied to stellar dynamics \cite{Bok,Mineur,Heggie,Heggie2},
and it is this second context that we have in mind here. ``Moon and Earth'' will become a
``star cluster'', and the role of the ``Sun'' will be played by the ``Galactic Center''.

Assuming, for simplicity, that the motion is in the plane, the $z$ coordinate can be dropped.
Then, for  approximately circular orbits, the first-order approximation to Newton's gravitational
equations provides us with  Hill's equations \cite{Gurfil,Heggie,ZGH},
\beq\begin{array}{lll}
m_a \bigl(\ddot{x}_a -2\omega \dot{y}_a-3\omega^2 x_a\Bigr)
&=&
\displaystyle\sum_{b\ne a} \frac{Gm_am_b(x_b-x_a)}{|\vx_a-\vx_b|^3}\,,
\\[8pt]
m_a \bigl(\ddot{y}_a + 2 \omega \dot{x}_a \bigr ) &=&
\displaystyle\sum_{b\ne a}\frac{Gm_am_b(y_b-y_a)}{|\vx_a-\vx_b|^3}\,
\end{array}
\label{Hilleqs}
\eeq
In these equations $\vx_a=(x_a,y_a)$ are the coordinates of star No $\underline{a}$
measured in a rotating coordinate system whose origin lies on the a Keplerian orbit with
$r=R,\, \theta=\omega t$. The $x$-axis is radial so that $r= R+x\,,$ and the $y$ axis is tangent to
the orbit.
$
\omega^2={GM}/{R^3}
$
[where $M$ is the mass of the Galaxy] is the angular velocity of a circular Keplerian
orbit with radius $R$. The origin $(x=y=0)$ of this frame represents hence the reference
orbit; our investigations  concern the behavior in its neighborhood.

The linear-in-velocity terms in (\ref{Hilleqs}) correspond to the Coriolis force induced
in a rotating coordinate system.

The motion of the ``Galactic Center'' is neglected.
The only remnant of the influence of the Galaxy on the star cluster corresponds, in the
first-order approximation, to the repulsive anisotropic harmonic term in the (radial)
$x$-equation, which arises from balancing attractive gravitational and repulsive centrifugal
forces.

The right hand sides represent the mutual gravitational interactions between the stars.

The Coriolis force plays a role analogous to a uniform magnetic field, turning our study
here analogous to the one in the Landau problem of the preceding section.

In the stellar context, a particularly interesting question is that of
\textit{ escape from the Galaxy} \cite{Heggie,Heggie2,Gurfil,Binney}. For individual stars
the answer is complicated, only allowing for a numerical treatment. The motion of the
center of mass (COM),
\beq
\vx=
\barray{c}x\\ y\earray=
\frac{\sum_{a}m_a\vx_a}{\sum_a m_a}\,,
\label{COM}
\eeq
is much simpler, though~: inter-stellar interactions drop out by Newton's third law,
leaving us with the simple planar system,
\beq
\begin{array}{llll}
\ddot{x}&-\,2\omega\dot{y}-3\omega^2x&=&0,
\\[2pt]
\ddot{y}&+\,2\omega\dot{x}&=&0.
\end{array} \,
\label{HillSymmeqs}
\eeq

These equations describe the oscillations of the center of mass of the considered star cluster
in the neighborhood of a reference Keplerian circle, represented here by $x=0,\, y=0$.

The
interest of studying the COM-problem is underlined by the fact that a cluster is formed by a huge number
of stars, --- in fact, of the order of a million \cite{Heggie2}.

\medskip
\subsection{Star escape~: Hall motions in the Sky}

For the COM, the problem of escape can also be reduced to that of a guiding center.

\begin{figure}
\begin{center}
\includegraphics[scale=1]{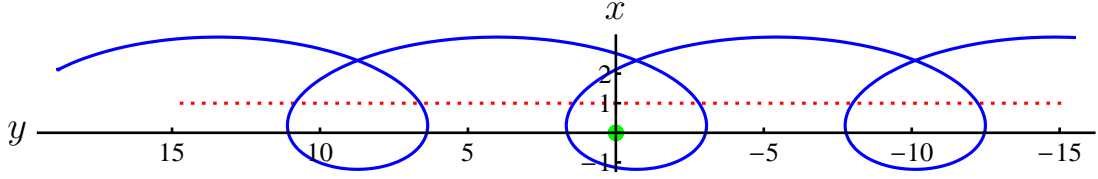}\quad
\vspace{-8mm}
\end{center}
\caption{{\it Trajectory of the center of mass in the Hill problem (in \blue{blue}) in the rotating
coordinates $x,y$. The \red{red} dotted straight horizontal line indicates the trajectory of the
guiding center about which the center of mass performs ``flattened elliptic motion''.
The heavy (\green{green}) dot in the origin stands for the reference Keplerian orbit. Note the unconventional orientation of the axes.}\\
}
\label{Hillplot}
\end{figure}

Equations (\ref{HillSymmeqs}) are readily solved  \cite{ZGH} as
\beq
\begin{array}{llll}
x(t)&=&\displaystyle\frac{A}{\omega}\sin\omega t -
\displaystyle\frac{B}{\omega}\cos\omega t
&+\;x_0,\,
\\[8pt]
y(t)&=&2\displaystyle\frac{A}{\omega}\cos\omega t+
2\displaystyle\frac{B}{\omega}\sin\omega t
&-\;\displaystyle\frac{3}{2}\omega tx_0+y_0.
\end{array}
\label{xysymm}
\eeq
where $A, B, x_0, y_0$ are integration constants. (\ref{xysymm}) is an ellipse centered at
$(x_0,y_0-\frac{3}{2}\omega x_0 t)$ with major axes lying along the $y$ direction, whose
centers drift along $y$ with constant speed $-\frac{3}{2}\omega x_0$, see Fig. \ref{Hillplot}.

$\bullet$ For initial condition $x_0=y_0=0$, the trajectory is an ellipse centered at the origin
and oriented along the $y$ direction,
\beq
\vX_+(t)=
\barray{c}X_+^1(t)\\ X_+^2(t)\earray=
\barray{c}
\displaystyle\frac{A}{\omega}\sin\omega t -
\displaystyle\frac{B}{\omega}\cos\omega t
\\[8pt]
2\displaystyle\frac{A}{\omega}\cos\omega t+
2\displaystyle\frac{B}{\omega}\sin\omega t
\earray\,.
\label{flatell}
\eeq

$\bullet$  For the particular, ``Hall'' initial conditions,
\beq
\barraynb{lllllll}
X_-^1(0)&=&x_0,&\qquad &X_-^2(0)&=&y_0,
\\
\dot{X}_-^1(0)&=&0,&\qquad &\dot{X}_-^2(0)&=&-\smallover3/2\omega x_0
\earraynb,\;
\eeq
we get, instead,
\beq
\vX_-(t)=\barray{c}
X_-^1(t)\\ X_-^2(t)\earray=
\barray{c}
x_0\\ -\frac{3}{2}\omega tx_0+y_0
\earray,
\label{Hallsol}
\eeq
identified with the motion the of \emph{guiding center}. The latter solution arises
when the harmonic and the inertial forces  \emph{cancel}, so that the COM drifts
with the \emph{Hall velocity} perpendicularly to the harmonic field,
\beq
\dot{X}_-^i=\varepsilon^{ij}\frac{E^j}{B},
\quad
\hbox{where}
\quad
eE^1=3\omega^2X_-^1,
\quad
E^2=0,
\quad
eB=2\omega.
\label{Halllawbis}
\eeq
i.e., it performs a \emph{Hall motion}.

The general solution, (\ref{xysymm}), is the sum of two particular solutions,
\beq
\vx(t)=\vX_+(t)+\vX_-(t).
\eeq
The coordinate  $\vX_+(t)$ describes, in particular, the relative motion about the guiding center. Note that an initial condidition $x(0)=0,\,y(0)=y_0\neq0$ yields shifted elliptic trajectories with the fixed point $x(t)=0,\,y(t)=y_0\neq0$ on the $y$-axis as guiding center.

\kikezd{Translation into fixed coordinates}

\begin{figure}
\begin{center}
\includegraphics[scale=.46]{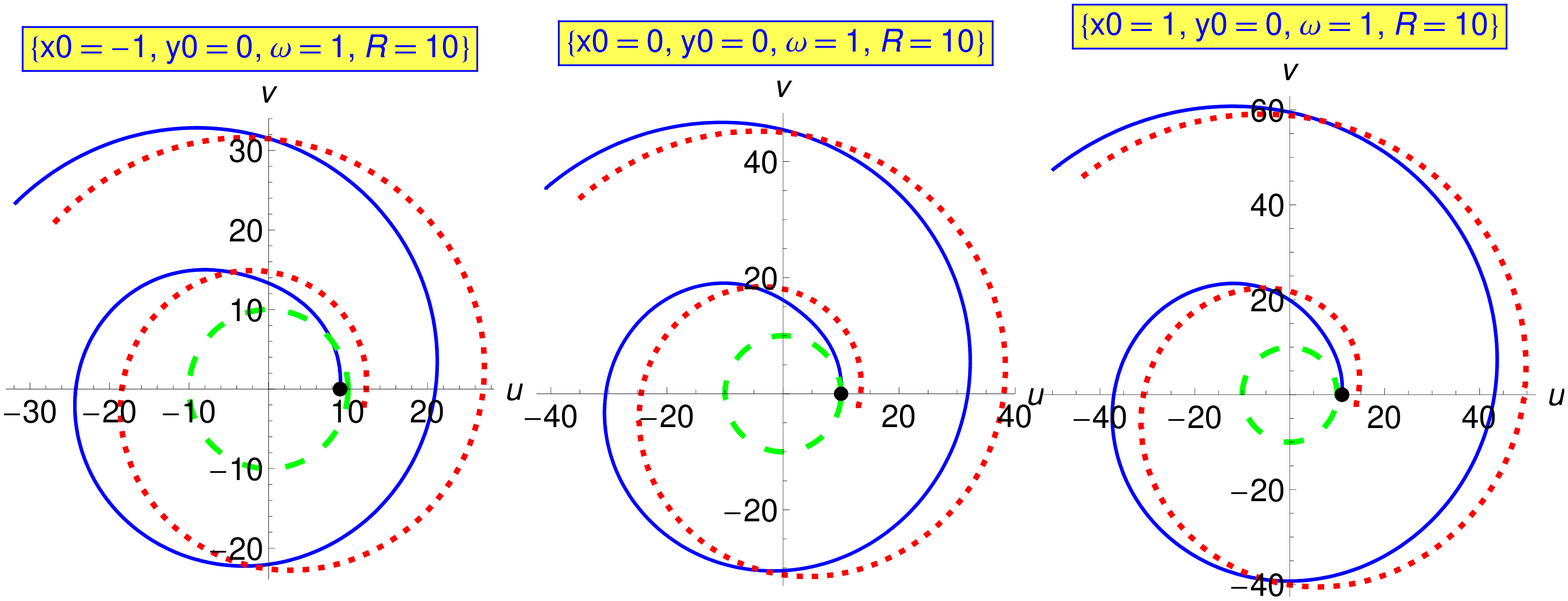}\\[5pt]
\includegraphics[scale=.44]{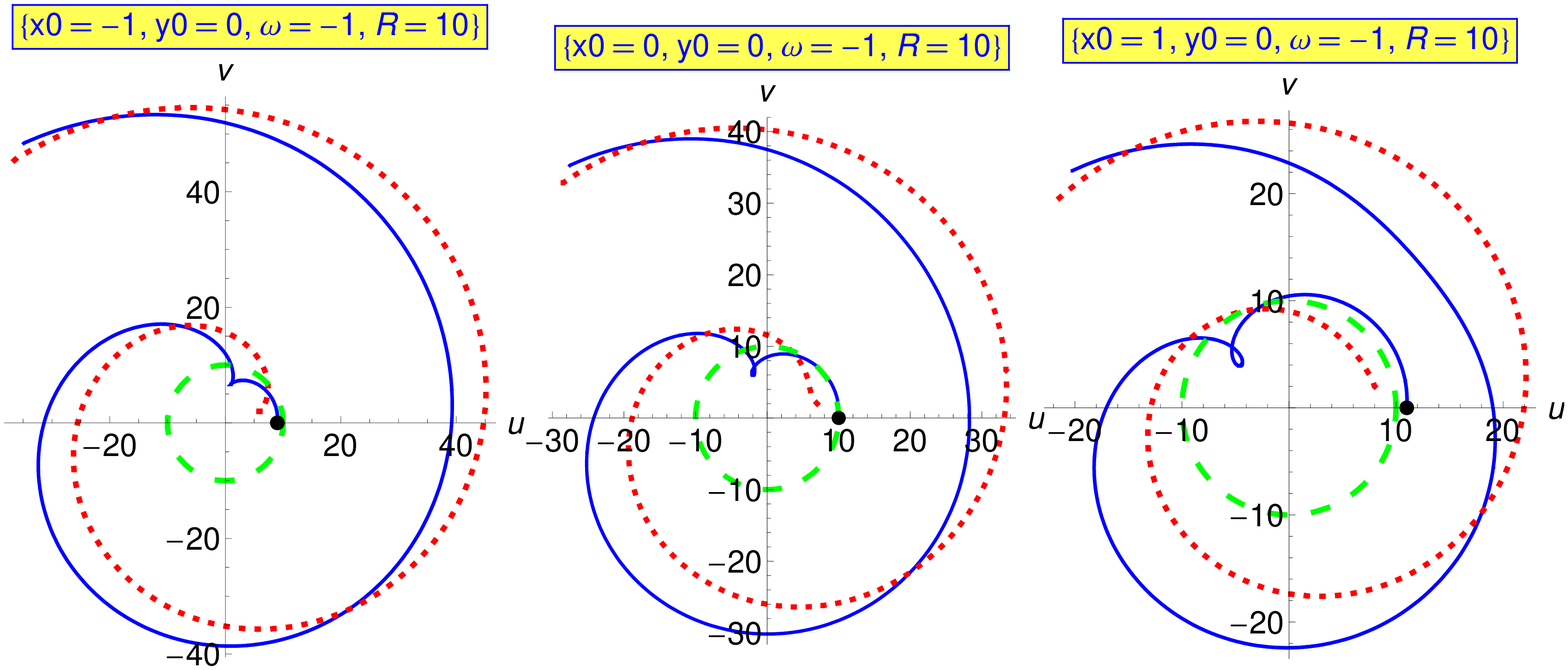}
\vspace{-10mm}
\end{center}
\caption{\it For all initial conditions $u_0=R+x_0$, with $x_0\neq 0,\,v_0=y_0$ the motions (in \blue{blue}) are unbounded: our star cluster escapes. The dotted \red{red} line is the guiding center, and the dashed \green{green} circle is the reference Keplerian trajectory, which corresponds to the heavy dot at the origin on Fig.~\ref{Hillplot}.}
\label{Efig}
\end{figure}

\begin{figure}
\begin{center}
\includegraphics[scale=.34]{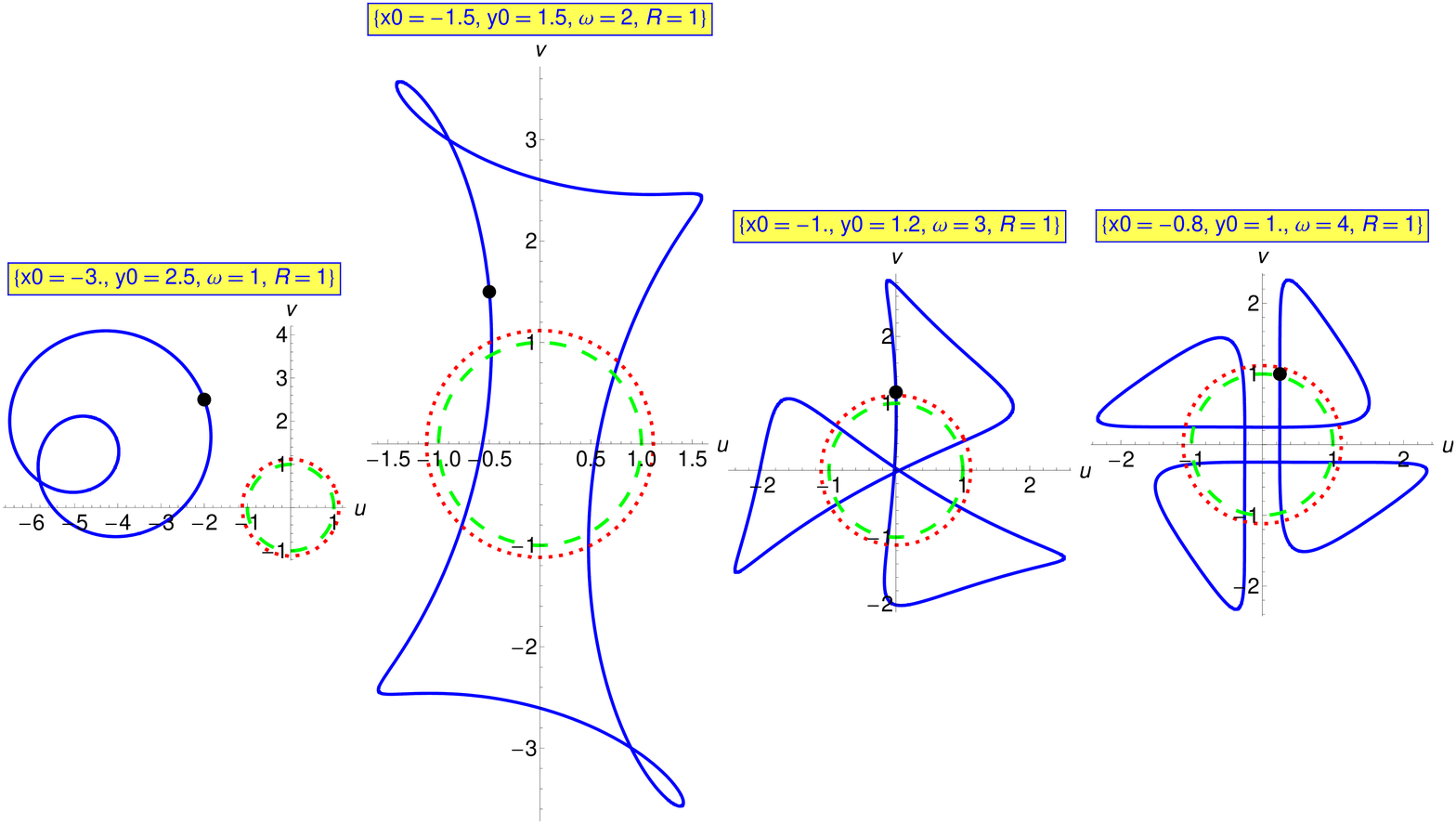}
\\[6pt]
\includegraphics[scale=.34]{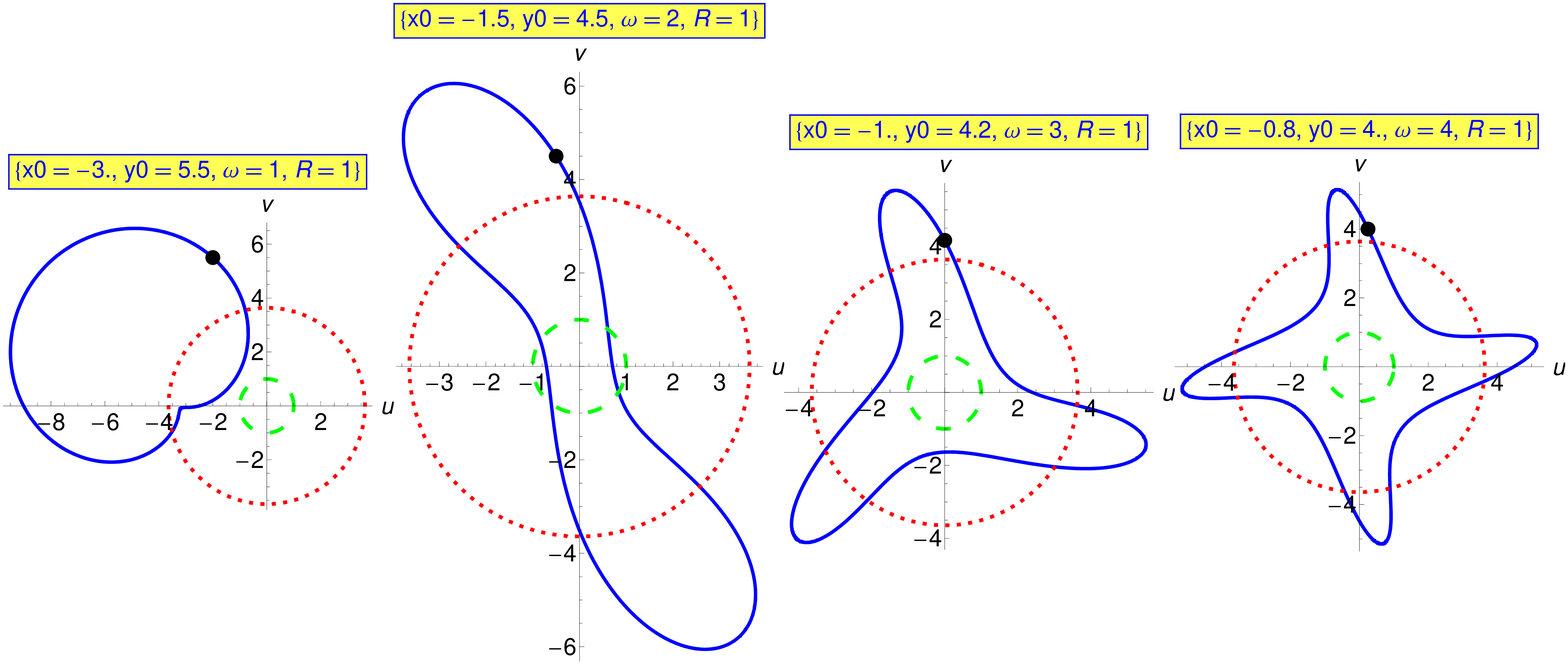}
\vspace{-7mm}
\end{center}
\caption{
{\it Bounded trajectories (in \blue{blue}) in the fixed reference frame $(u,v)$
arise for initial condition $u_0=R$ i.e. $x_0=0$ (only),  indicated by a dot.
 The shape of the trajectory strongly
depends on the initial condition $y_0$ and on the choice of the Keplerian circle (in \green{green}
).
The \red{red} circle is the guiding center.} }
\label{Bfig}
\end{figure}

So far we worked in a rotating coordinate system. Putting
\beq\barraynb{lll}
u(t)&=&\big(R+x(t)\big)\cos\omega t-y(t)\sin\omega t,
\\
v(t) &=& \big(R+x(t)\big)\sin\omega t+y(t)\cos \omega t
\,,
\earraynb
\eeq
allows us to express our results in the fixed coordinate system $(u,v)$, as shown on
Figs.~\ref{Efig}  and \ref{Bfig}. As said before, our Keplerian reference circle corresponds to the origin of the $x-y$ plane,
$x(t)=y(t)=0$.

Our clue now is that the \emph{motion is only bounded when that of the guiding center is}.
We focus therefore our attention to $\vX_-(t)$. By eqn. (\ref{Hallsol}) the guiding centers
move forcelessly, governed by the Hall law. Then, although the coordinate $x(t)$ remains
bounded for any choice of the parameters, motion in the \emph{tangential} ($y$) direction
\emph{increases} the distance from the galactic center in all cases. \emph{All motions with
initial condition $x_0\neq0$ [i.e., $u_0\neq R$] are, therefore unbounded}.

Bounded motions \emph{only} arise for initial condition on the $y$-axis $x_0=0$, when, in
the co-moving frame, $(x,y)=(0,y_0)$ is fixed. Then the expansion is stopped and the
guiding center trajectory is in fact a Keplerian circle. As a result of the oscillatory
motion of $\vX_+(t)$ the final trajectories are however, quite complicated, as illustrated on
Fig. \ref{Bfig}.

The distinction between bounded  and unbounded motions corresponds to the negative semi-definite nature of the
effective potential energy in the co-moving frame,
\beq
V_{eff}=-3m\omega^2x^2,
\eeq
for which only $x=0$ is a neutral direction. For all initial conditions with $x_0\neq0$
the repulsive harmonic potential is indeed of the tidal nature, implying that all such
solutions are unstable.


Looking at Figs. \ref{Efig}, it is tempting to think at those celebrated spiral arms of galaxies.
One should keep in mind, however, that our investigations are only valid in the neighborhood
of the Keplerian orbit characterized by $x=y=0$. For large values of $x$ and $y$ our linear
approximation breaks down, and it would plainly be abusive to draw any conclusion about the
long-range behavior from studying the linearized equations (\ref{HillSymmeqs}). Our linear
approximation is only justified when all coordinates $x_a,y_a$ are small when compared to the
Keplerian radius $R$. The tendency to escape is, however, reflected by their initial behavior
studied here.

\newpage
\subsection{Newton-Hooke-type symmetry in the Hill problem}

Having understood the importance of the center-of-mass decomposition, we turn to study the
symmetry which makes it work -- namely that of Newton-Hooke with no rotations.

The coordinates $\vX_{\pm}$ introduced above can, indeed, be completed to chiral coordinates,
namely by putting \cite{ZGH}
\beq
\vx=\vX_{+}+\vX_{-},
\qquad
{p}^1=\frac{1}{2}\omega X_{+}^2,
\qquad
{p}^2=-2\omega X_{+}^1-\displaystyle\frac{3}{2}\omega X_{-}^1.
\eeq

Our investigations can in fact be generalized to exotic  particles i.e. to non-commutative
particles, see \cite{ZH-Hill}. Skipping details, we state that, in all cases one finds that
ordinary translations and certain ``time dependent translations'' (also called ``generalized boosts''),
\beq\barraynb{lll}
\vPi&=&
\barray{c}
X_-^1(t)
\\[4pt]
X_-^2(t)+\displaystyle\frac{3}{2}\omega tX_-^1(t)
\earray,
\\[28pt]
\vK&=&\barray{c}X_+^1(t)\cos(\omega/\Delta) t-
\smallover{1}/{2\Gamma}X_+^2(t)\sin(\omega/\Delta) t
\\[8pt]
 2\Gamma X_+^1(t)\sin(\omega/\Delta) t
+X_+^2(t)\cos(\omega/\Delta) t
\earray,
\earraynb
\label{exoHillNH}
\eeq
are conserved, where
\beq
\Delta=1-2m\omega\theta,
\qquad
\Gamma= 1-3m\theta\omega/2.
\eeq
Their commutation relations are, once again, those of two exotic Heisenberg algebras with
central charges $-(2/m\omega)$ and  $(\Gamma/\Delta )(2m\omega)$, respectively,
\beqa
\{\Pi^1,\Pi^2\}=-\frac{2}{m\omega}\,,
\label{XHalg}
\qquad
\{K^1, K^2\}=\frac{\Gamma}{\Delta}\,\frac{2}{m\omega}\,,
\qquad
\{\Pi_i, K_j\}=0\,.
\eeqa
In the commutative case $\theta=0$ so that $\Gamma=\Delta=1$, and the one-parameter centrally
extended symmetry found in \cite{ZGH} is recovered. The Hamiltonian,
\beq
H=H_++H_-=\displaystyle\frac{m\omega^2}{2}\left(
X_{+}^1X_{+}^1
+\frac{1}{4\Gamma^2}X_{+}^2X_{+}^2
\right)
-
\displaystyle\frac{3m\omega^2}{8}
X_{-}^1X_{-}^1,\quad
\label{HillHam}
\eeq
is also conserved. Its commutation relations with translations and boosts read
\beqa\barraynb{lllllll}
\{H,\Pi^1\}&=&0,
\quad&
&\{H,\Pi^2\}&=&\displaystyle\frac{3}{2}\omega\,\Pi^1,
\\[8pt]
\{H,K^1\}&=&-\displaystyle\frac{\omega^*}{2\Gamma}K^2,
\quad&
&\{H,K^2\}&=&2\Gamma\omega^*\,K^1\,.
\earraynb
\label{Hamcomm}
\eeqa

As rotational symmetry is plainly broken, the total symmetry of the Hill problem is
\emph{Newton-Hooke \emph{without} rotations}.

\section{Anisotropic harmonic oscillator}
\label{anisotropic}

We note that the Hill problem is in fact a maximally anisotropic ``one sided'' oscillator.
The case of a  general anisotropic oscillator is worth studying in some detail therefore.

\subsection{Chiral coordinates}\label{chiralcoord}

Consistently with the general theory sketched in Section \ref{Landau}, an ``exotic''
[i.e., non-commutative] charged  harmonic oscillator in the plane in a homogenous
magnetic field $B$ is described by the symplectic form and Hamiltonian,
\beqa
\Omega&=&
d{p}^i\wedge dx^i+\frac{\theta}{2}\varepsilon^{ij}d{p}^i\wedge d{p}^j+\frac{eB}{2}
\varepsilon^{ij}dx^i\wedge dx^j,
\label{exosymp}
\\[8pt]
H&=&\frac{\vp^2}{2m}+V,
\qquad
V=\half k_1x_1^2+\half k_2x_2^2,
\label{osciham}
\eeqa
respectively, with the parameters having the same physical interpretation as before.
The spring constants $k_1$ and $k_2$ may or may not be identical.

The idea of Alvarez \textit{et al}. \cite{AGKP} has been to \emph{combine} chiral oscillators.
Multiplying both the symplectic form and the Hamiltonian (or, alternatively, the Lagrangian)
by the same overall constant $\mu$,
\beq
\Omega\to\mu\,\Omega,
\quad
H\to\mu\, H
\qquad\hbox{i.e.}\qquad
L\to \mu\, L\,,
\nn
\eeq
would not change the equations of motion. But what happens, if we multiply them with
\emph{different} coefficients before \emph{adding} them~? Conversely, can we decompose
a given system into two chiral parts~? To answer these questions we introduce, following
\cite{AGKP,ZH-chiral,ZGH}, new coordinates on the phase space,
\beqa
x^i=X_{+}^i+X_{-}^i,
\qquad
p^1=\alpha_{+}X_{+}^2+\alpha_{-}X_{-}^2,
\quad
p^2=-\beta_{+}X_{+}^1-\beta_{-}X_{-}^1,
\label{chico}
\eeqa
where the coefficients $\alpha_\pm$ and $\beta_\pm$ will be determined from the requirement
that both the symplectic form and the Hamiltonian should split into two uncoupled
one-dimensional subsystems we shall call chiral components. Inserting (\ref{chico}) into
(\ref{exosymp}) shows that the symplectic form splits as $\Omega=\Omega_++\Omega_-$, whenever
\begin{equation}
\alpha_{-}+\beta _{+}
-\theta\alpha_-\beta_+=eB,
\qquad
\alpha_{+}+\beta_{-}
-\theta\alpha_+\beta_-=eB.
\end{equation}
Similarly, inserting (\ref{chico}) into (\ref{osciham}) yields that the Hamiltonian splits as
$H=H_++H_-$ when
\begin{equation}
\alpha_{+}\alpha_{-}+mk_2=0,
\qquad
\beta_{+}\beta_{-}+mk_1=0.
\end{equation}
Then a tedious calculation allows choosing
\begin{eqnarray}
\alpha_{+}&=&-\frac 1{2(eB+\theta mk_1)}\Bigg(
-e^2B^2+m(k_2-k_1)+\theta^2m^2k_1k_2
\label{aplus}
\\[6pt]
&&+\sqrt{4mk_1(eB+\theta mk_2)^2
+\big(e^2B^2-m(k_1-k_2)-\theta^2m^2k_1k_2\big)^2}
\Bigg)\,,\nn
\end{eqnarray}
\begin{eqnarray}
\alpha_{-}&=&\frac 1{2(eB+\theta mk_1)}\Bigg(
e^2B^2-m(k_2-k_1)-\theta^2m^2k_1k_2
\label{aminus}
\\[6pt]
&&+\sqrt{4mk_1(eB+\theta mk_2)^2+\big(e^2B^2-m(k_1-k_2)-\theta^2m^2k_1k_2\big)^2}
\Bigg)\,,\nn
\end{eqnarray}
and
\begin{eqnarray}
\beta_{+}&=&-\frac 1{2(eB+\theta mk_2)}\Bigg(
-e^2B^2+m(k_1-k_2)+\theta^2m^2k_1k_2
\label{bplus}
\\[6pt]
&&+\sqrt{4mk_1(eB+\theta mk_2)^2+\big(e^2B^2-m(k_1-k_2)-\theta^2m^2k_1k_2\big)^2}
\Bigg)\,,
\nn
\end{eqnarray}
\begin{eqnarray}
\beta_{-}&=&\frac 1{2(eB+\theta mk_2)}\Bigg(
e^2B^2-m(k_1-k_2)-\theta^2m^2k_1k_2
\\[6pt]
&&+\sqrt{4mk_1(eB+\theta mk_2)^2+\big(e^2B^2-m(k_1-k_2)-\theta^2m^2k_1k_2\big)^2}\Bigg)\,.
\nn
\end{eqnarray}
provides us with decomposed symplectic form and the Hamiltonian,
\begin{eqnarray}
\Omega  &=&\Omega_++\Omega_-=
\label{Omegadecomp}
\\[8pt]
&&\underbrace{\left(-\alpha_{+}-\beta_{+}+\theta\alpha_{+}\beta
_{+}+eB\right)}_{\mu_+}dX_{+}^1\wedge dX_{+}^2
+
\underbrace{\left(-\alpha_{-}-\beta_{-}+\theta\alpha_{-}\beta _{-}+eB\right)}_{\mu_-}
dX_{-}^1\wedge dX_{-}^2\,,\qquad\nn
\eeqa
and
\beqa
H &=&H_++H_-=\frac{1}{2m}\times
\label{Hdecomp}
\\[8pt]
&&\Big[\left(\beta_{+}^2+mk_1\right) X_{+}^1X_{+}^1+\left(
\alpha_{+}^2+mk_2\right)X_{+}^2X_{+}^2+\left( \beta _{-}^2+mk_1\right)
X_{-}^1X_{-}^1+\left( \alpha _{-}^2+mk_2\right) X_{-}^2X_{-}^2\Big],\qquad\nn
\end{eqnarray}
respectively.

\begin{itemize}
\item
For $\theta=0$ the commutative cases \cite{AGKP, ZGH} are recovered;

\item
For $k_1=-3m\omega^2$, $k_2=0$ and $B=2\omega$, we get
\beq\barraynb{llll}
\alpha_+=0,
\qquad
&\alpha_-=\displaystyle\frac{m\omega}{2\Gamma},
\quad
&\beta_+=\displaystyle\frac{3}{2}m\omega,
\qquad
&\beta_-=2m\omega,
\\[12pt]
\mu_+=\displaystyle\frac{m\omega}{2},
\qquad
&\mu_-=-\displaystyle\frac{\Delta}{\Gamma}\frac{m\omega}{2},
&&
\earraynb
\eeq
and the results found before in the Hill Problem \cite{ZGH,ZH-Hill} are obtained [up to  interchanging $\vX_+$ and $\vX_-$];

\item
When $k_1=k_2$ our oscillator is \emph{isotropic}. Then $\alpha_\pm=\beta_\pm$, and
(\ref{Omegadecomp}-\ref{Hdecomp}) reduce to  the chiral decomposition
for the [exotic] Landau problem with harmonic force, studied in \cite{ZH-chiral};

\item
For $k_1=k_2=0$ the oscillator is switched off, and the system reduces to the
purely-magnetic non-commutative Landau problem  \cite{NCLandau,AGKP,ZH-chiral}.

\end{itemize}

\subsection{Motions}

Let us assume that none of the coefficients $\mu_\pm$ vanishes. Then it follows from
(\ref{Hdecomp}) that our chiral coordinates satisfy the Poisson bracket relations
\beq
\big\{X_+^i,X_+^j\big\}=-\frac{1}{\mu_+}\,
\varepsilon^{ij},
\qquad
\big\{X_+^i,X_-^j\big\}=0
\qquad
\big\{X_-^i,X_-^j\big\}=-\frac{1}{\mu_-}\,
\varepsilon^{ij}\,.
\label{chircommrel}
\eeq
The equations of motion read therefore
\beq\barraynb{llll}
m\mu_\pm\dot{X}_\pm^1&=&-&\big(\alpha_\pm^2+mk_2)X_\pm^2,
\\[6pt]
m\mu_\pm\dot{X}_\pm^2&=&\;&\big(\beta_\pm^2+mk_1)X_\pm^1.
\earraynb
\label{X+chireqmot}
\eeq

Both chiral components $X_\pm$ are governed, hence, by uncoupled equations which are
reminiscent of those of $1d$ harmonic oscillators, to which they reduce, however,
only in the isotropic case, $k_1=k_2$.

\begin{figure}
\begin{center}
\includegraphics[scale=.5]{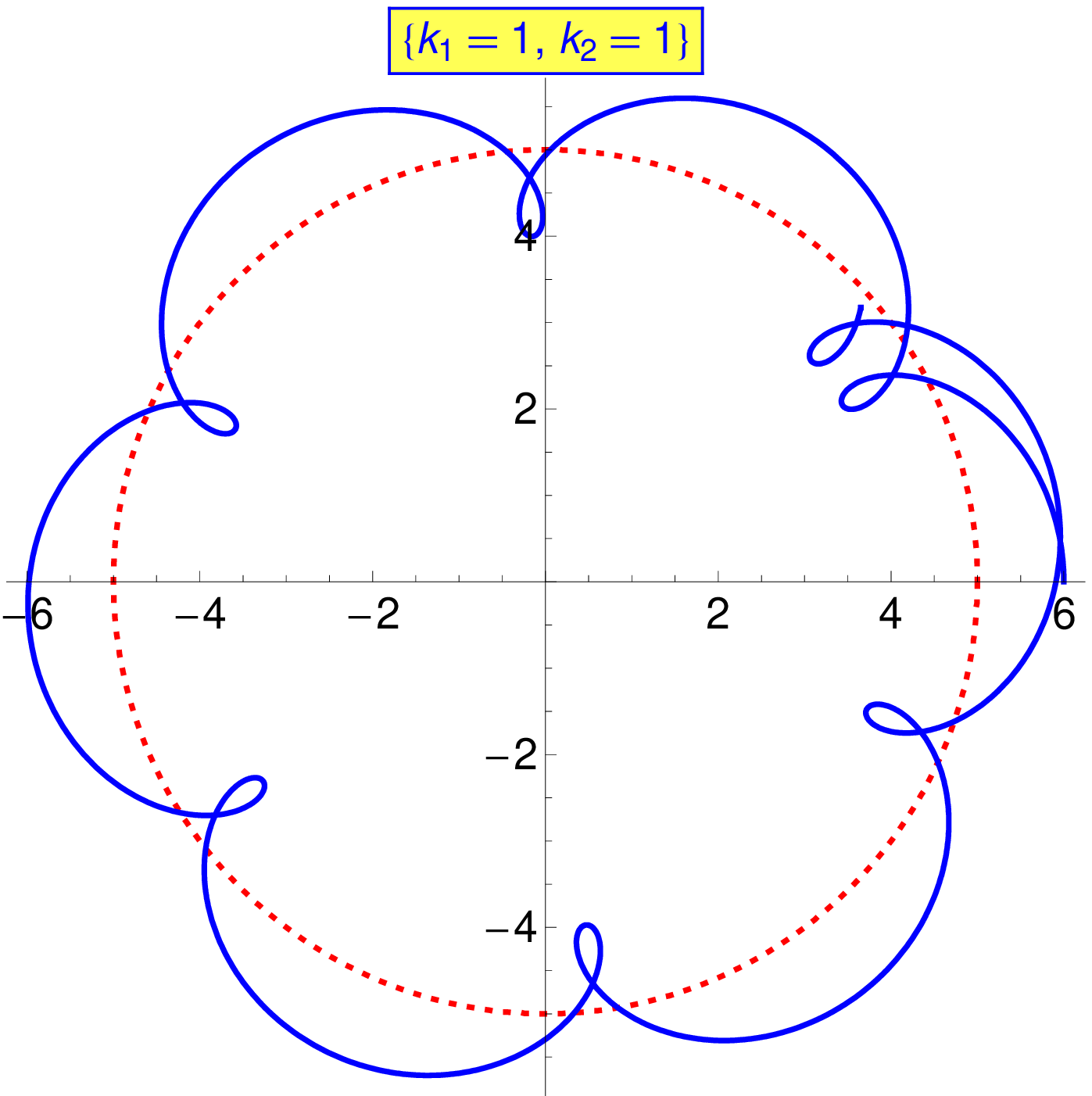}
\;
\includegraphics[scale=.6]{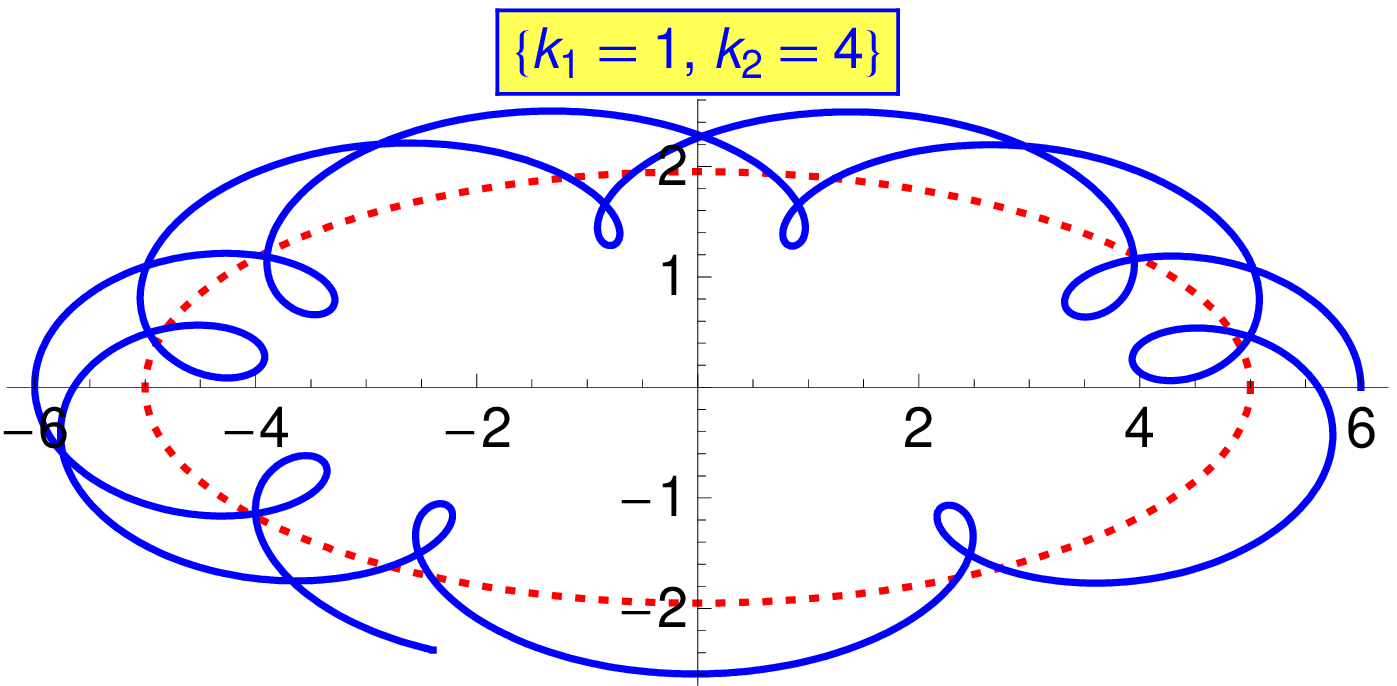}
\vspace{-8mm}
\end{center}
\caption{\it ``Epicyclic'' motion of an oscillator in a constant magnetic field.
(a) is the isotropic case $k_1=k_2$, and (b) is an anisotropic case with $k_1=4k_2$.
The \blue{blue} line is the physical trajectory, $\vx = \vX_{+} + \vX_{-}$, and the dotted
\red{red} line is the  motion of the guiding center, $\vX_{+}(t)$.
}
\label{aniosci}
\end{figure}

Assuming $\alpha_\pm^2+mk_2\neq0$ \footnote{In the Hill case $\omega_+=0$ and the
$\vX_+$-dynamics is free, while $\omega_-=\omega/\Delta$, cf. \cite{ZH-Hill}.},
eqns. (\ref{X+chireqmot}) are solved by
\beq\barraynb{llll}
X_\pm^1&=&\quad A_\pm\cos\omega_\pm t+B_\pm\sin\omega_\pm t,\qquad&
\\[6pt]
X_\pm^2&=&F_\pm\,
\Big(A_\pm\sin\omega_\pm t-B_\pm\cos\omega_\pm t
\Big),
\qquad\qquad
&F_\pm=\displaystyle\sqrt{\frac{\beta_\pm^2+mk_1}{\alpha_\pm^2+mk_2}}\,,
\earraynb
\label{eqmotsol}
\eeq
where the frequencies read
\beq
\omega_\pm=\frac{\sqrt{(\alpha_\pm^2+mk_2)(\beta_\pm^2+mk_1)}}{m\mu_\pm}\,.
\eeq
Both $\vX_\pm$-trajectories are ellipses, as illustrated in Fig. \ref{aniosci}. Note that
the frequencies, $\omega_+$ and $\omega_-$ are in general different even in the isotropic case,
and the curves do not close therefore.

\subsection{Symmetries}\label{chiralsymm}

Eqns. (\ref{eqmotsol}) allow us to infer that
\beq\barraynb{lll}
A_\pm&=&X_\pm^1\cos\omega_\pm t+
\displaystyle\frac{1}{F_\pm}X_\pm^2\sin\omega_\pm t,
\\[8pt]
B_\pm&=&
X_\pm^1\sin\omega_\pm t-
\displaystyle\frac{1}{F_\pm}X_\pm^2\cos\omega_\pm t
\earraynb
\label{eqmotsol2}
\eeq
are conserved. A direct calculation yields, furthermore, for both labels $\pm$, the
uncoupled Heisenberg algebra relations
\beq
\{A_\pm,B_\pm\}=-\frac{1}{F_\pm\mu_{\pm}},
\qquad
\Big\{(\,\cdot\,)_+\,,\,(\,\cdot\,)_-\Big\}=0.
\label{genHalg}
\eeq
Adding the Hamiltonian (\ref{Hdecomp}), the doubly-centrally-extended rotation-less
Newton-Hooke algebra is obtained.

Both sets of chiral coordinates $\vX_\pm$ describe $2d$ symplectic vectorspaces.
The symplectic forms $\Omega_\pm$ are plainly symmetric under phase-space chiral rotations,
$
\vX_\pm\to R(\vX_\pm).
$
None of the Hamiltonians $H_\pm$ is symmetric in general, though. The natural diagonal action,
\beq
\vx=\vX_++\vX_-\to R(\vX_+)+R(\vX_-)=R(\vx),
\eeq
is \emph{not} a symmetry therefore~: \emph{rotations are broken by the anisotropy}.

In the \emph{isotropic case},
\beq
k_1=k_2,
\eeq
however, we have $\alpha_\pm=\beta_\pm$ and  the coefficients of the quadratic terms both
in $H_+$ and $H_-$ are hence identical, so that the chiral rotations
$
\vX_\pm\to R(\vX_\pm)
$
do act as symmetries for the components~:  \emph{rotational symmetry is  restored}.
The square-root factors  in (\ref{eqmotsol}) become unity, $F_\pm=1$, and the trajectories
become circles. The frequencies,
\beq
\omega_\pm=\frac{\alpha_\pm^2+mk}{m\mu_\pm},
\eeq
are not identical, though, since $\alpha_+\neq \alpha_-$ and $\mu_+\neq \mu_-$
in general, cf. (\ref{aplus}) -- (\ref{aminus}) and (\ref{Omegadecomp}).

It is worth recording that, in terms of chiral coordinates, the total angular momentum,
(\ref{totangmom}), is also  decomposed, as
\beq
J=J_++J_-\,,
\qquad
J_{+}=
\frac{eB}{2}\left(\vec{X}_{+}\right)^2,
\qquad
J_{-}=-\Delta\,\frac{eB}{2}\left(\vec{X}_{-}\right)^2,
\label{chiralangmom}
\eeq
where $\Delta=1-eB\theta $, as before. Its conservation, $\dot{J}=0$, can also be checked
directly, using the equations of motion.

We just mention that the singular case $\mu_+=0$ or $\mu_-=0$, leading to Hall-type motion,
can be  dealt with as in the previous occasions, \cite{DHexo,ZH-Kohn-II,NCLandau,ZH-chiral,ZH-Hill}.

\goodbreak
\section{Systems with prescribed NH-type symmetry}\label{orbitconst}

Any physical theory consists of some mathematical structure together with a set of operational
rules relating the abstract notions entering this structure to physically measurable quantities.
It happens quite often that various theories share the same mathematics and differ only in
their interpretation. It is, therefore, interesting to analyse the mathematical theories
commonly appearing in different physical contexts. In most cases (if not all, at least as
far as the basic microscopic theories are concerned) the choice of the dynamical equations
is based on symmetry principles. If the symmetry group is selected the form of dynamics is
strongly restricted; in some cases all admissible dynamics can be even fully classified.
Once this is done one may look for theories sharing the same formal structure but differing
in the operational meaning of formal notions used.

In the present paper we are interested in physical systems exhibiting Newton-Hooke type
symmetry. This section is devoted to the classification and analysis of formal properties
of dynamical systems possessing such type of symmetry. We wish to show that the results
of the previous Sections, which provide physical examples of the systems under consideration,
fit, in fact, into some general framework. To this end, we assume that the dynamics under
consideration is invariant under the transitive action of some Lie group $G$, and then
classify all such symplectic manifolds upon which $G$ acts by symplectically. The proper
tool for doing this is provided by the \emph{orbit method} \cite{SSD, Kirillov,Arnold}.

Our choice for the group $G$ is dictated by the following considerations. As far as
possible, we would like to allow for a generalization which includes both the Galilei
and the Newton-Hooke groups, and also the ``rotation-free part'' of the latter. The main
characteristic features are therefore the following:

(i) there exists generators (namely of boosts and momenta) which, via the orbit method,
yield the basic canonical variables;

(ii) The Hamiltonian equations of motion are linear in the latter variables; the Hamiltonian
belongs therefore to the Lie algebra itself, and acts linearly on the remaining variables.

We want our generalization to be a minimal one in that no further symmetry generators
beyond the above ones should be included. Such generators will  appear later however
for specific values of the structure constants.

Guided by these considerations, we start  with the following Lie algebra commutation
relations,
\beq\barraynb{lllll}
[\xi_i,\xi_j]&=&i\omega_{ij}M,
&\qquad
&i,j=1,\dots, 2N
\\[6pt]
[M,\xi_i]&=&0,&&
\\[6pt]
[M,H]&=&0,&&
\\[6pt]
[H,\xi_i]&=&iA_{ij}\xi_j,&&
\earraynb
\label{algrel1}
\eeq
where $\omega=(\omega_{ij})$ is a non-singular antisymmetric matrix. The only non-trivial
Jacobi identity,
\beq
\left[\big[H,\xi_i\big],\xi_j\right]+\hbox{(cyclic)}=0,
\eeq
yields the constraint
$
A_{ik}\omega_{kj}-A_{jk}\omega_{ki}=0,
$
i.e., that
$
B=A\omega
$
is a symmetric matrix, $B^T=B$.

The algebra (\ref{algrel1}) admits the Casimir operator of the form
\beq
C=MH-\half X_{ij}\xi_i\xi_j,
\label{Casimir}
\eeq
where without loss of generality we can assume that $X=(X_{ij})$ is symmetric.
$C$ commutes with all generators, provided
$
A=-\omega X.
$

Collecting our results, our algebra reads
\beq\barraynb{lll}
[\xi_i,\xi_j]&=&i\omega_{ij}M,
\\[6pt]
[M,(\,\cdot\,)]&=&0,
\\[6pt]
[H,\xi_i]&=&-i\omega_{ik}X_{kj}\xi_j,
\\[6pt]
C&=&MH-\half X_{ij}\xi_i\xi_j,
\earraynb
\label{algrel2}
\eeq
and is uniquely defined by choosing the non-singular antisymmetric matrix $\omega$
and the symmetric matrix $X$.

The next step is to classify the inequivalent algebras (\ref{algrel2}).
Under the invertible transformation
\beq
{\xi_i}'=D_{ij}\xi_j,
\qquad
\hbox{det }\, (D_{ij})\neq0,
\eeq
The matrices $\omega$ and $X$ transform according to
\beq
\omega'=D\omega D^T, \qquad
X'=(D^{-1})^TXD^{-1}\,.
\label{Deq}
\eeq
Using the latter we can find the ``canonical'' form in any class of equivalent
algebras (\ref{algrel2}).

In what follows we shall restrict ourselves to the case $2N=4$, the generalization
to arbitrary $N$ being straightforward.

To complete our classification scheme some further assumptions on the matrix $X$
have to be made. The existence of the Casimir operator $C$ implies that, on each
orbit, the Hamiltonian is a quadratic function of the basic canonical variables,
to which a trivial term, representing the internal energy, has been added (see Appendix B).
Whether the energy is positive definite or not depends, therefore, on the choice of $X$.

The following cases will be considered separately.

\subsection{$X$ Positive definite}

Consider first the case of a positive definite matrix  $X$. By an appropriate choice of $D$
in eqns. (\ref{Deq}),   $X=I$ can be achieved. In fact, $X$, being symmetric, can be
diagonalized by a suitable orthogonal transformation. Then  an additional diagonal
transformation reduces $X$ to the unit matrix.

Assuming $X=I$, we still have some residual transformations left at our disposition. Namely,
as it is seen from eqns. (\ref{Deq}), $D$ can be taken to be an arbitrary orthogonal matrix,
without spoiling the condition $X=I$. The question is now to classify all antisymmetric
$4\times4$ matrices $\omega$ up to an orthogonal transformation. This problem is solved in
Appendix A (which is actually the Euclidean version of the classification problem for
electromagnetic field configurations under the action of the Lorentz group). As shown
in Appendix A, $\omega$ can be put into the form
\beq
\omega=\barray{cc}
0&\barraynb{cc}
\Omega_1&0
\\
0&\Omega_2
\earraynb
\\
\barraynb{cc}
-\Omega_1&0
\\
0&-\Omega_2
\earraynb
&0
\earray,
\qquad
\Omega_{1,2}>0.
\label{omegaform}
\eeq
Defining
\beq
B_1=\Omega_1^{-1}\xi_1,
\quad
B_2=\Omega_2^{-1}\xi_2,
\quad
P_1=\xi_3,
\quad
P_2=\xi_4,
\label{BPdef}
\eeq
one finds the following non-trivial commutators~:
\beq\barraynb{lll}
[B_i,P_k]=i\delta_{ik}M,
\\[4pt]
[H,B_i]=-iP_i,
\\[4pt]
[H,P_i]= i\Omega_i^2B_i,
\earraynb
\label{algrel3}
\eeq
together with
\beq
C=MH-\half\big(P_1^2+P_2^2+\Omega_1^2B_1^2
+\Omega_2^2B_2^2\big).
\label{Casi2}
\eeq
\kikezd{Orbits}

We can now apply the orbit method (Appendix B). Consider the coadjoint orbit parametrized
by $m>0$, the coordinate in dual space corresponding to the Casimir operator $M$ and
interpreted as the \emph{mass}, and by $\epsilon m$, corresponding to the Casimir
operator $C$ and interpreted as the \emph{internal energy}. Let $p_i,\,b_i,\,h,\, i=1,2$
be the relevant coordinates in the space dual to the Lie algebra (\ref{algrel3}). As shown
in Appendix B, the points of the coadjoint orbit are parametrized by $p_i$ and $b_i$.
Defining
\beq
q_i=b_i/m,
\label{defq}
\eeq
we find
\beqa
\big\{q_i,p_k\big\}=\delta_{ik},
\qquad
h=\left(\frac{p_1^2}{2m}+
\frac{m\Omega_1^2}{2}q_1^2\right)
+\left(\frac{p_2^2}{2m}+
\frac{m\Omega_2^2}{2}q_2^2\right)
+\epsilon.
\label{defq1}
\eeqa
Hence, we arrive at an in general \emph{anisotropic oscillator},
as the most general system with positive definite energy, admitting the symmetry
defined by the rotation-less Newton-Hooke commutation relations (\ref{algrel1}).

\subsection{$X$ semi-positive}

Let us consider the case when the matrix $X$ is semidefinite. We restrict ourselves to $X$
having a single zero eigenvalue (as in the Hill case). Then one can select the matrix $D$
in (\ref{Deq}) in such a way that $X$ acquires the form
\beq
X=\barray{cc}
\barraynb{cc}
1&0\\0&0\earraynb
&0
\\[6pt]
0&\barraynb{cc}
1&0\\0&1\earraynb
\earray\,.
\label{semiX}
\eeq
One can show again (see Appendix A) that the residual freedom in the choice of the
basis of our algebra allows us to put $\omega$ into the form (\ref{omegaform}).
Using again eqns (\ref{BPdef}), we find therefore
\beq\barraynb{lll}
[H,B_i]&=&-iP_i,
\\[4pt]
[H,P_1]&=& i\Omega_1^2B_1,
\qquad
[H,P_2]=0,
\\[4pt]
C&=&MH-\half\big(P_1^2+P_2^2+\omega_1^2B_1^2\big)
\earraynb
\label{algrel4}
\eeq
The orbit method yields, in this case, the dynamics describing a \emph{harmonic oscillator
in one direction, and free motion in the second one} --- as in the Hill problem \cite{ZGH}.

The case of multiple null eigenvalues of $X$ can be dealt with similarly.

\subsection{$X$ indefinite}

Let us drop, finally, the assumption of positive (semi)definiteness of $X$. We consider
in more detail the cases of two positive -- one negative -- one null eigenvalues. By an
appropriate choice of $D$ one can achieve
\beq
X=\barray{cc}
0&0
\\[2pt]
0&G\earray,
\qquad\hbox{where}\qquad
G=\hbox{diag}(-1,1,1).
\label{indefX}
\eeq
According to the results in Appendix A, the symplectic form $\omega$ can acquire three
canonical forms, namely those presented in  eqns (\ref{A5}) - (\ref{A18}) - (\ref{A19}).
Then the orbit method gives the following dynamical systems~:\\
(i)
\beqa
\label{dynsys1}
\big\{q_i,p_k\big\}&=&\delta_{ik},
\\[8pt]
h&=&\frac{p_1^2}{2m}+\left(\frac{p_2^2}{2m}
-\frac{m\Omega_2^2}{2}q_2^2\right)
+\epsilon\,;
\eeqa
(ii)
\beqa
\big\{q_i,q_j\big\}&=&\sigma\epsilon_{ij},
\qquad
\big\{p_i,p_j\big\}=\tau\epsilon_{ij},
\\[8pt]
h&=&\frac{p_1^2}{2m}+\left(\frac{p_2^2}{2m}
-\frac{q_2^2}{2}\right)
+\epsilon\,;
\eeqa
(iii)
\beqa
\big\{q_i,q_j\big\}&=&\sigma\epsilon_{ij},
\qquad
\big\{q_2,p_2\big\}=1,
\qquad
\big\{p_i,p_j\big\}=\tau\epsilon_{ij},
\\[8pt]
h&=&\frac{p_1^2}{2m}+\left(\frac{p_2^2}{2m}
-\tau^2\frac{q_2^2}{2m}\right)
+\epsilon\,.
\label{dynsys2}
\eeqa
The parameter $\sigma$ here is a clear indication of \emph{non-commuting nature} of
the coordinates $q_1$ and $q_2$ \footnote{In fact, $\sigma=\theta/(1-\theta\tau)$,
$\theta=\sigma/(1+\sigma\tau)$, where $\theta$ is the non-commutativity parameter.}.

The case of non(semi)definite Hamiltonian is the most involved one. Unlike in the
previous cases, after the ``canonical" Hamiltonian is fixed, there still remain
three inequivalent forms of the basic Poisson brackets.

The reason for that is clearly seen from the derivation given in Appendix A. The
$3\times 3$ submatrix $\omega_g$ of the matrix $\omega $ transforms, under the
transformations leaving the form of the Hamiltonian invariant, as an $\Ort(2,1)$
antisymmetric tensor. Its canonical form depends therefore on the value of the
``electromagnetic" invariant
\beq
\sum_{i=1}^{2} (\omega _{0i})^2 - (\omega _{12})^2.
\eeq
Depending on its value, the basic Poisson brackets can take different, inequivalent
forms (assuming the form of Hamiltonian is fixed). The labeling of variables in
equations (\ref{dynsys1}) - (\ref{dynsys2}) is dictated by our preference for
the form of the Hamiltonian, rather then that of the Poisson brackets. It must
be stressed, however, that the final choice of appropriate variables should be
dictated by additional assumptions, not resulting from symmetry considerations only.

\par
As an example, let us consider the planar Hill equations, as presented in Refs.
\cite{ZGH,ZH-Hill}. The Hamiltonian reads
\beqa
H=\frac{1}{2m}\big(p_1^2+p_2^2\big)-\frac{3m\omega^2}{2}q_2^2,
\eeqa
and yields Hill's equations for the following Poisson brackets,
\beqa
\big\{q_i,p_j\big\} = \delta_{ij},
\qquad
\big\{p_i,p_j\big\} = 2m\omega\epsilon_{ij}\,,
\eeqa
where the commutative case, $\sigma=0$, has been chosen for simplicity. The other
parameter is $\tau^2=3m^2\omega^2$, and $B=2m\omega$ is the effective ``magnetic'' field.
Let us put
\beqa
\xi_1 = \lambda\, q_1, \qquad \xi_2 = \sqrt{3m}\omega\, q_2,
\qquad
\xi_3 = \frac{p_1}{\sqrt{m}},
\qquad
\xi_4 =\frac{p_2}{\sqrt{m}}
\eeqa
with $\lambda \neq 0 $ arbitrary. Then $H$ acquires the standard form
\beqa
H = \frac{1}{2} \big( \xi _3^2 + \xi _4^2 - \xi _2^2 \big),
\eeqa
and the relevant Poisson brackets read
\beqa
\big\{\xi _2,\xi_4 \big\} = \sqrt{3}\,\omega,
\qquad
\big\{\xi_2,\xi _4 \big\} &=& 2\omega.
\eeqa
Therefore one finds, with the notations of Appendix A,
\beq
\omega_{01} =0,
\qquad
\omega_{02} = \sqrt{3}\,\omega,
\qquad
\omega_{12} = 2\omega,
\qquad
{\vec \omega}^2 - \omega_{12}^2 = -\omega^2 < 0.
\eeq
According to the classification given in Appendix A, we are dealing with
the case (\ref{A16}), and the ``canonical" form of the Poisson brackets is
given by eqn.  (\ref{A18}), in full agreement with the results of Refs.
\cite{ZGH,ZH-Hill}.

\subsection{Additional symmetries}\label{further}

We now study the question of additional symmetries. Consider the case of a
positive definite Hamiltonian. As it has been shown in the previous Section,
the initial algebra can be put into the form
\beqa
[\xi_i,\xi_j]&=&i\omega_{ij}M,
\\[4pt]
[H,\xi_i]&=&-i\omega_{ij}\xi_j,
\\[4pt]
[M,\,\cdot\,]&=&0,
\eeqa
where $\omega$ is given by eqn. (\ref{omegaform}). We add a new generator
$J$ which is assumed to obey
\beq
[J,M]=0,\qquad
[J,H]=0,\qquad
[J,\xi_i]=ij_{ik}\xi_k,
\eeq
where $j=(j_{ik})$ is an appropriate matrix. The two additional Jacobi identities
\beqa
[J,[H,\xi_i]]+(\hbox{cycl})=0
\qquad
[\xi_i,[J,\xi_j]]+(\hbox{cycl})=0
\label{Jacobi2}
\eeqa
yield
$
j\omega+\omega j^T=0,
\,
j\omega-\omega j=0.
$
Hence $j=-j^T$, and the general solution reads

\vskip2mm
(i) $\Omega_1\neq\Omega_2$,
\beq
j=\alpha\barray{cc}
0&\barraynb{cc}
1&0\\ 0&0\earraynb
\\[4pt]
\barraynb{cc}
-1&0\\ 0&0\earraynb
&0
\earray
+\beta\barray{cc}
0&\barraynb{cc}
0&0\\ 0&1\earraynb
\\[4pt]
\barraynb{cc}
0&0\\ 0&-1\earraynb
&0
\earray\,,
\label{j1}
\eeq
(ii) $\Omega_1=\Omega_2$,
\beq
j=\alpha\barray{cc}
0&\barraynb{cc}
1&0\\ 0&0\earraynb
\\[4pt]
\barraynb{cc}
-1&0\\ 0&0\earraynb
&0
\earray
+\beta\barray{cc}
0&\barraynb{cc}
0&0\\ 0&1\earraynb
\\[4pt]
\barraynb{cc}
0&0\\ 0&-1\earraynb
&0
\earray
+
\gamma\barray{cc}
0&\barraynb{cc}
0&1\\ 1&0\earraynb
\\[4pt]
\barraynb{cc}
0&-1\\ -1&0\earraynb
&0
\earray
+\delta\barray{cc}
\barraynb{cc}
0&1\\-1&0\earraynb
&0
\\[4pt]
0&
\barraynb{cc}
0&1\\ -1&0\earraynb
\earray\,.
\label{j2}
\eeq

Before explaining the meaning of the particular solutions, let us note that,
once the equations (\ref{Jacobi2}) are obeyed, there exists a second Casimir
operator, namely
\beq
\widetilde{C}=MJ-\half Y_{ij}\xi_i\xi_j,
\label{Casibis}
\eeq
where $Y=-\omega^{-1}j$.
If the coadjoint orbit is parametrized by the value $m\tilde{\sigma}$ of the Casimir
$\widetilde{C}$, eqn. (\ref{Casibis}) \footnote{$m$ is the eigenvalue of the
operator $M$.} yields the expression for $J$ which, as in the case of the Hamiltonian,
consists of the sum of the quadratic term plus  the ``internal'' contribution to $J$,
\beq
mJ=\half Y_{ij} z_i z_j+m\tilde{\sigma},
\label{mJ}
\eeq
where the $z$'s are the basic variables parametrizing the points on the orbit
(cf. (\ref{B1}) in Appendix B). Eqn. (\ref{mJ}) allows us interpret $\tilde{\sigma}$
as the \emph{internal angular momentum}, analogous to the \emph{internal energy},
$\epsilon$, in our previous formul{\ae}, cf. \cite{SSD}. Using the general solution
for $j$, eqns. (\ref{j1}) and (\ref{j2}), one finds the following generators as
functions on the coadjoint orbit (up to an internal part)~:

(i) for $\Omega_1\neq\Omega_2$,
\beq
H_1=\frac{p_1^2}{2m}+
\frac{m\Omega_1^2}{2}q_1^2,
\qquad
H_2=\frac{p_2^2}{2m}+
\frac{m\Omega_2^2}{2}q_2^2\,;
\eeq
(ii) for $\Omega_1=\Omega_2=\Omega$,
\beqa
\barraynb{lllllll}
H_1&=&\displaystyle\frac{p_1^2}{2m}+
\displaystyle\frac{m\Omega^2}{2}q_1^2,
&\qquad
&H_2&=& \displaystyle\frac{p_2^2}{2m}+
\displaystyle\frac{m\Omega^2}{2}q_2^2,
\\[12pt]
J&=&q_1p_2-q_2p_1,
&\qquad
&Z&=&p_1p_2+
m^2\Omega^2q_1q_2.
\earraynb
\label{4integrals}
\eeqa
\par
The meaning of the above expressions is clear. First of all, for an anisotropic
oscillator we have two integrals, corresponding to the partial energies; our system
is integrable.

For equal frequencies, the dynamics is superintegrable~: there are three functionally
independent integrals. One can choose the angular momentum as the third one. The four
integrals in eqn. (\ref{4integrals}) are linearly independent but they are functionally
dependent. Note also that our integrals (\ref{4integrals}), being quadratic in canonical
variables, form the u$(2)$ Lie algebra --- the well-known dynamical algebra of a
two-dimensional isotropic harmonic oscillator. In fact, if one defines
\beqa
V_{1} \equiv \frac{1}{2} J,
\qquad V_2 \equiv \frac{1}{2\Omega }(H_2-H_1),
\qquad
V_3 \equiv \frac{1}{2m\Omega} (p_1 p_2 + m^2 \Omega^2 q_1 q_2),
\eeqa
the resulting Poisson brackets algebra reads
\beqa
\big\{V_i, V_j\big\} = \epsilon_{ijk}V_k,
\eeqa
i.e., span the $\su(2)$ algebra. The fourth generator, namely the Hamiltonian,
\beq
V_0 \equiv H,
\eeq
can also be added \cite{AGKP}. $V_0$ commutes with all other $V$'s, completing the
$\su(2)$ algebra into the unitary algebra u$(2)$.

Let us remark that even for $\Omega_1\neq\Omega_2$ there exists an additional integral,
provided the ratio of the frequencies is rational,
\beq
r=\Omega_1/\Omega_2=m/n.
\eeq
It is, however, no longer quadratic in the canonical variables, yielding a $W$-algebra,
instead of a Lie algebra \cite{aniOsci}. In fact, an additional integral of the motion
which yields our system superintegrable can be constructed as follows. One defines the
classical counterparts of the creation/annihilation operators by
\beq
a_i=q_i-\frac{ip_i}{m\Omega_i},\,
\qquad
\bar{a}_i=q_i+\frac{ip_i}{m\Omega_i}\,.
\eeq
It is then easy to check that
\beq
C^{\,n,m} = (a_1)^n(a_2)^m
\eeq
is an \emph{integral of the motion}. In the isotropic case $n=m=1$, for example,
\beq
C\equiv C^{1,1}=\frac{Z}{m^2\omega^2}+\frac{i}{m\Omega}J
\eeq
is a combination of those conserved quantities in the second line of (\ref{4integrals}),
namely of the angular momentum and the ``mixed'' quantity denoted by $Z$.

The integral $C^{n,m}$ is functionally independent of the partial energies, $H_{1,2}$.
Moreover, there are no further independent (and explicitly time-independent) integrals;
therefore, the Poisson bracket between $H_{1,2}$ and $C^{n,m}$ are functionally
expressible in terms of them, and form a finite $W$-algebra  \cite{Walg}.

\vskip3mm
Let us conclude this section with some remarks. We have shown, at least in the case
of (semi)definite hamiltonian, that there exists a unique ``canonical" form of the
underlying dynamics. However, the choice of this canonical form is dictated by
mathematical simplicity rather than by physical requirements which are, in fact,
additional assumptions. It seems reasonable to assume, generally, that the physical
variables are those which convert the system into (non-commutative) anisotropic
oscillator in a uniform magnetic background. This can be always done because our
canonical form may be converted back into any other hamiltonian form obeying the symmetry
assumptions. Therefore, we end up with Theorem 1, as stated in the Introduction.

\section{The Bargmann point of view}\label{Bargmann}

The NH symmetry of an isotropic oscillator can conveniently be derived by ``importing''
the Galilei symmetry of a free particle using Niederer's transformation, which maps
every half period of the oscillator onto a free particle  \cite{NiedererOsc,ZH-Kohn}.
One way of seeing this is to work within Duval's ``Bargmann'' framework, where
classical non-relativistic motions are null geodesics of a suitable relativistic
spacetime \cite{Eisen,DGH}. Null geodesics are invariant w.r.t. conformal transformations,
and Niederer's transformation,
\beq
T=\frac{\tan \omega t}\omega\, ,
\qquad
\vec{X}=\frac{\vec{x}}{\cos \omega t}\,,
\qquad
S=s-\frac{\omega r^2}2\tan \omega t
\label{Ntrafo}
\eeq
maps indeed every half oscillator period conformally onto the space-time which describes
a free particle,
\beq
d\vX^2+2dTdS =
\frac{1}{\cos^2\omega t}
\left(d\vx^2+2dtds-\omega^2r^2dt^2\right).
\eeq
This trick can \emph{not} work for an anisotropic oscillator, though, otherwise
the latter would also carry a full NH symmetry including rotation.

An anisotropic oscillator is described by the metric \footnote{The Bargmann space
(\ref{aniBarg}) is not conformally flat as its Weyl tensor does not vanish,
unless $\omega_1=\omega_2$.}
\beq
d\vx^2+2dtds-\big(\omega_1^2x_1^2+\omega_2^2x_2^2\big)dt^2.
\label{aniBarg}
\eeq
Applying Niederer's transformation (\ref{Ntrafo}) i.e.
\beq
t=\frac{\arctan \omega T}\omega\,,
\quad
\vx=\frac{\vX}{\sqrt{1+\omega
^2T^2}},
\quad
s=S+\frac 12\frac{\omega^2\vX^2T}{1+\omega^2T^2}
\eeq
 with some $\omega$ then yields
\begin{eqnarray*}
\frac{1}{1+\omega^2T^2}\left(d\vX^2+2dTdS-\frac{\omega_1^2-\omega^2}
{\left(1+\omega ^2T^2\right)^2}X_1^2dT^2-\frac{\omega_2^2-\omega^2}
{\left(1+\omega^2T^2\right)^2}X_2^2dT^2\right) .
\end{eqnarray*}
Now choosing either $\omega=\omega_1$ or $\omega=\omega_2$ eliminates \emph{one},
but not \emph{both} oscillator terms, leaving us with
\begin{eqnarray}
d\bar{s}^2&=& \frac{1}{1+\omega_2^2T^2}\left(d\vX^2+2dTdS-\frac{\omega_1^2-\omega_2^2}
{\left(1+\omega_2^2T^2\right)^2}X_1^2dT^2\right)\nn
\\[8pt]
&=&\frac{1}{1+\omega_1^2T^2}\left(
d\vX^2+2dTdS-\frac{\omega_2^2-\omega_1^2}{\left(1+\omega_2^2T^2\right)^2}X_2^2dT^2
\right).
\label{onesidedmetric}
\end{eqnarray}
[where we should have put indices $1$ or $2$ on $\vX$, depending on our choice of $\omega$].
For both choices we get, hence, a maximally anisotropic ``one-sided'' ``Hill-type''
system, with Newton-Hooke symmetry --- \emph{except} in the isotropic case
\beq
\omega_1=\omega_2,
\eeq
when \emph{both} oscillator terms drop out, leaving us with a free system carrying its
\emph{full Galilei symmetry}. The latter can then be ``re-imported'' through the inverse
of the Niederer transformation (\ref{Ntrafo}) to yield full Newton-Hooke symmetry.

In conclusion, the ``prototype system'' is of the ``Hill type'', with its rotation-less
Newton-Hook symmetry --- which, in the isotropic case, degenerates to a free particle
with restored rotational symmetry.

\section{Conclusion}\label{Concl}

Souriau \cite{SSD} attributes the center-of-mass decomposition of a \emph{free}
non-relativistic system to \emph{Galilei symmetry}, more precisely, to an invariant
Abelian subgroup of it, whose  existence is rooted in turn in the cohomology of
the Galilei group \cite{SSD}. Remarkably, it is this same cohomology which rules
central extensions \cite{LL}.

In this paper we performed an analogous study in the \emph{Landau problem}, based on  the
Newton-Hooke group. The clue is that  Newton-Hooke and Galilean symmetries are indeed
``hiddenly the same'' \cite{ZH-Kohn}, and have therefore identical cohomological
structures \cite{NHcoho}.

The intuitive content of Kohn's theorem, i.e., the relation between [Newton-Hooke] symmetry
and center-of-mass, is now clear~: each particle, taken individually, would carry such a
symmetry; Kohn's condition is precisely what is needed to extend this symmetry to the
center-of-mass, which will hence represent the motion of all particles collectively.

A method for finding  approximate solutions of the 3-body problem of Celestial
Mechanics, also used in Galactic Dynamics \cite{Heggie}, is referred to as the
\emph{Hill Problem}. The latter also has a  symmetry reminiscent of the Newton-Hooke
one, except for rotations, which are missing.

In Section\ref{HillSection} we applied, for the first time in our knowledge,
the center-of-mass decomposition to study of the star escape problem in Hill's
framework. But as mentioned above, the very possibility of such a decomposition
relies on the existence of an invariant Abelian subgroup of the symmetry group
(which can either be the Galilei group or the rotation-less Newton-Hooke group).
Our main result here is to find, conversely, the most general mechanical system
with the latter symmetry, namely the anisotropic harmonic oscillator in a uniform
magnetic background.

At the technical level, the Hill Problem is a particular case of an anisotropic
harmonic oscillator in an effective magnetic field.

In this paper, we performed a similar study for a \emph{general anisotropic harmonic oscillator}.

All our investigations here have been purely classical. The decomposition of Newton-Hooke
symmetry into Heisenberg algebras is, however, particularly useful for the quantum
description, see \cite{Sochi,ZH-chiral} for details.

\vskip5mm
\appendix{\bf Appendix A}

\renewcommand\thesection{A}
\numberwithin{equation}{section}

We find here the canonical form of the $4\times4$ antisymmetric nonsingular
matrix $\omega$ undergoing the transformation
\beq
\omega\to D\omega D^T,
\quad\hbox{where $D$ obeys}\quad
D^TXD=X,
\label{A1-2}
\eeq
$X$ being the matrix defined in eqn. (\ref{Casimir}).

As it has been noted in the main text, $X$ can be put into canonical form,
which depends on the assumption concerning the eigenvalues of $X$.

Consider first $X$ positive definite; then we can put $X=I$. As a result
$D$ is orthogonal and we have to find the canonical form of $\omega$ under $\ort(4)$
transformations (\ref{A1-2}). This resembles the problem of classifying the
electromagnetic field configurations
under the Lorentz group $\Ort(3,1)$. Guided by this analogy, we define
\beq
f_i=\omega_{0i},\qquad
g_i=\half\epsilon_{jk}\omega_{jk}.
\label{A3}
\eeq
Note that $f_i$ and $g_i$ transform like vectors under ${\rm SO}(3)$ transformations
acting on the last three coordinates. Moreover,
det $\omega\sim (\vec{f}\cdot\vec{g})^2$,
so that $\vec{f}\cdot\vec{g}\neq0$, i.e.,
$\vec{f}\neq0$, $\vec{g}\neq0$ and $\vec{f}$
is not perpendicular to $\vec{g}$.

Let us consider the rotation in the plane spanned by the $O$-axis, and the
axis which is orthogonal to it and defined by the unit vector $\vec n$.
The transformation rules under such a rotation read
\beqa
\vec{f}_{\parallel}' &=&\vec f_{\parallel},
\qquad
\vec {f}_{\perp}' = \vec f_{\perp } \cos \varphi  + (\vec n \times \vec g_{\perp }) \sin \varphi,
\nonumber
\\
\vec {g}_{\parallel}' &=&\vec g_{\parallel},
\qquad
\vec {g}_{\perp}' = \vec g_{\perp } \cos \varphi  - (\vec n \times \vec f_{\perp }) \sin \varphi,
\label{A4}
\eeqa
where $\parallel (\perp )$ denotes the component parallel (orthogonal) to $\vec n$.
If $\vec f \nparallel \vec g$
we put
\beq
\vec n = \frac{\vec f \times \vec g}{|\vec f \times \vec g|} \qquad\hbox{and}\qquad
\sin{2\varphi} = \frac{2|\vec f \times \vec g|}{\vec f ^2 + \vec g ^2}
\eeq
 to achieve $\vec f \parallel \vec g$.
Then by $SO(3)$ rotation one gets further $f_i =\Omega _1 \delta _{i2}$,
$\Omega_1 > 0$, $g_i = - \Omega _2 \delta _{i2}$.
Renumbering, if necessary, $1\leftrightarrow 3$ (which is an $O(3)$ transformation)
we let $\Omega _2 > 0$. Due to definition (\ref{A3}),
\beq
\omega =\barray{cccc}
0&f_1&f_2&f_3\\-f_1&0&g_3&-g_2\\
-f_2&-g_3&0&g_1\\
-f_3&g_2&-g_1&0
\earray\, =
\barray{cccc}
0&0&\Omega _1&0\\
0&0&0&\Omega _2\\
-\Omega _1&0&0&0\\
0&-\Omega _2&0&0
\earray\,.
\label{A5}
\eeq

Consider next the case of semidefinite $X$ with one null eigenvalue. Then $X$
can be put in the form
$
X=\barray{cc}
0&0\\0&I_3
\earray\,.
$
Put
\beq
D = \barray{cc}
d&A\\
B&U
\earray .
\label{A7}
\eeq
Eqns. (\ref{A1-2}) implies $B=0$, $U\in \Ort(3)$, so $D$ acquires the form
$
D=\barray{cc}
d&A\\
0&U
\earray
, \, d \neq 0.
$
 Then, with $\omega_{ij}=\epsilon_{ijk}g_k$,
\beq
D\omega D^T =\barray{cc}
0&dfU^T + A\omega _gU^T\\
-dUf^T + U\omega_gA^T&U\omega_g U^T
\earray.
\label{A9}
\eeq
Here $\omega _g $ is an antisymmetric matrix, so it belongs to the algebra $\so(3)$.
One can choose therefore
$U\in \SO(3)$ such that
\beq
U\omega _g U^T = \barray{ccc}
0&0&\Omega _2\\
0&0&0\\
-\Omega _2&0&0
\earray
, \qquad
\Omega _2 > 0.
\label{A10}
\eeq

Consider now the elements $dfU^T+A\omega_gU^T=dfU^T+AU^TU\omega_gU^T$. Call
\beq
dfU^T\equiv (\tilde f_1, \tilde f_2, \tilde f_3), \qquad AU^T\equiv (\tilde a_1,
\tilde a_2, \tilde a_3).
\label{A11}
\eeq
Then
\beqa
dfU^T + AU^TU\omega _gU^T &=& (\tilde f_1,\tilde f_2,\tilde f_3) +
(\tilde a_1,\tilde a_2,\tilde a_3)
\barray{ccc}
0&0&\Omega _2\\
0&0&0\\
-\Omega _2&0&0
\earray\nn
\\[12pt]
&=& (\tilde f_1, \tilde f_2, \tilde f_3) + (-\Omega _2\tilde a_3, 0, \Omega _2\tilde a_1).
\label{A12}
\eeqa
Knowing $f$, $U$ and $d$ one determines $\tilde f_{1,2,3}$ and chooses
$\tilde a_{1,3}$ in such a way that
\beq
dfU^T+AU^TU\omega_gU^T = (0, \tilde f_2, 0),
\quad \tilde f_2\neq 0.
\label{A13}
\eeq

By an appropriate choice of $d$ we get $0<\tilde f_2\equiv \Omega_1$; so
(\ref{A9}) acquires the form (\ref{A5}).

Finally, let $X$ have two positive, one negative and one zero eigenvalue.
Without loosing generality, we put

\beq
X=\barray{cc}
0&0\\
0&G
\earray
, \qquad G\equiv diag(-1, 1, 1)
\label{A14}
\eeq
With $D$ of the form (\ref{A7}) eqn. (\ref{A1-2}) yields $B = 0$, $U \in \Ort(2,1)$ ;
$D\omega D^{T}$ has the same form (\ref{A9}).

Consider now
$U\omega_{g}U^{T}$. Again proceeding along the same lines as in the
classification of electromagnetic field configurations, we find that
$U\omega _{g}U^{T}$ can acquire three ``canonical" forms:
\beq
U\omega _{g}U^{T}=\barray{ccc}
0&0&\Omega _2\\
0&0&0\\
-\Omega _2&0&0
\earray
, \qquad \Omega_2 > O
\label{A15}
\eeq

\beq
U\omega _{g}U^{T}=\barray{ccc}
0&0&0\\
0&0&\Delta \\
0&-\Delta &0
\earray
, \qquad  \Delta > O
\label{A16}
\eeq

\beq
U\omega _{g}U^{T}=\barray{ccc}
0&0&\Sigma \\
0&0&\Sigma \\
-\Sigma &-\Sigma &0
\earray
, \qquad \Sigma  \neq O\,.
\label{A17}
\eeq

If (\ref{A15}) holds the same reasoning as previously leads to eqn.
(\ref{A5}). In the second case
 \beq
\omega =\barray{cccc}
0&\Omega &0&0\\
-\Omega &0&0&0\\
0&0&0&\Delta\\
0&0&-\Delta &0
\earray
, \qquad \Omega > O ,\qquad \Delta > 0\,.
\label{A18}
\eeq

Finally, if \ref{A17} holds,
\beq
\omega =\barray{cccc}
0&\Omega &0&0\\
-\Omega &0&0&\Sigma \\
0&0&0&\Sigma\\
0&-\Sigma&-\Sigma &0
\earray \,.
\label{A19}
\eeq

\vskip5mm
\appendix{\bf Appendix B}

\renewcommand{\theequation}{B\thesection.\arabic{equation}}
\renewcommand
\appendix{\appendix}{\setcounter{equation}{0}}

\par
We consider here the orbit method for the Lie algebra (\ref{algrel1}).
The general element of the dual space can be written as
\beq
h\tilde H + m\tilde M + z_i\tilde \xi ^i.
\label{B1}
\eeq
Consider the coadjoint action of $g = \exp{(iy^k\xi _k)}$. It reads
\beqa
m' &=& m,\nonumber
\\
z_i ' &=& z_i + \omega _{ki} y^k m,
\\
\label{B2}
h' &=& h + y^k \omega_{kl} X_{lj} z_j + \frac{1}{2} y^k y^l \omega_{lm}
\omega_{kj} X_{mj} m.\nonumber
\eeqa

Assuming $m \neq 0$ and using the fact that $\omega $ is invertible, we
conclude that each orbit contains
the points corresponding to $z_i = 0$. The set of these points forms the
coadjoint orbit of the stability subgroup
of the relations $z_i = 0$. However, the latter is generated by $M$ and $H$,
so the coadjoint orbits are trivial.
We conclude that $z_i = 0$ define exactly one point on coadjoint orbit.
Therefore, generating the whole orbit by
the action of our group on that point we conclude that the orbit can be
parametrized by the variables $z_i$ and
\beq
h = \epsilon + \frac{1}{2m} X_{ij} z_i z_j,
\label{B3}
\eeq
where $\epsilon $ is the value of $h$ at the point $z_i = 0$ (internal energy).
The basic Poisson bracket reads
\beq
\big\{ z_i, z_j \big\} = \omega _{ij} m,
\label{B4}
\eeq
which completes the description.

The additional symmetry generators can be dealt with in a similar way.

In the case of two degrees of freedom and (semi)definite $H$ it is convenient
to identify the ``physical" generators
as described by eqns. (\ref{BPdef}) (i.e. to single out the boosts and momenta).
In this basis the counterparts of dual
coordinates $z_i$ are denoted by $p_i$ and $b_i$ (cf. eqns.  (\ref{defq})
and (\ref{defq1})).

\begin{acknowledgments}
We would like to thank J. Balog, J. Binney, B. Erdi, G. Gibbons,
 D. Heggie and M. Plyushchay for  their interest and for  correspondence.
P.A.H  acknowledges  hospitality at the \textit{
Institute of Modern Physics} of the Lanzhou branch of
the Chinese Academy of Sciences.
 This work was  partially supported by the National Natural Science Foundation of
China (Grant No. 11035006 and 11175215) and by the Chinese Academy of Sciences
visiting
professorship for senior international scientists (Grant No. 2010TIJ06).
K.A., J.G and P.K. have been supported by the Polish Ministry of Science (grant N
N202 331139).
\end{acknowledgments}


\end{document}